\documentclass[]{iopart}

\expandafter\let\csname equation*\endcsname\relax
\expandafter\let\csname endequation*\endcsname\relax
\usepackage{amsmath}
\usepackage{graphicx}
\usepackage{siunitx}
\usepackage{glossaries}
\usepackage{hyperref}
\usepackage[table]{xcolor}

\newacronym{LISA}{LISA}{Laser Interferometer Space Antenna}
\newacronym{LPF}{LPF}{LISA Pathfinder}
\newacronym{ESA}{ESA}{European Space Agency}
\newacronym[longplural={massive black hole binaries}]{MBHB}{MBHB}{massive black hole binary}
\newacronym[longplural={verification galactic binaries}]{VGB}{VGB}{verification galactic binary}
\newacronym[longplural={galactic binaries}]{GB}{GB}{galactic binary}
\newacronym{EMRI}{EMRI}{extreme-mass ratio inspiral}

\newacronym{GW}{GW}{gravitational wave}
\newacronym{SGW}{SGW}{stochastic gravitational wave}
\newacronym{SGWB}{SGWB}{stochastic gravitational-wave background}
\newacronym{GR}{GR}{general relativity}

\newacronym{INReP}{INReP}{initial noise reduction pipeline}
\newacronym{TDI}{TDI}{time-delay interferometry}
\newacronym{MOSA}{MOSA}{movable optical sub-assembly}
\newacronym{OMS}{OMS}{optical metrology system}
\newacronym[longplural={test masses}]{TM}{TM}{test mass}
\newacronym{LDC}{LDC}{LISA Data Challenge}
\newacronym{FOM}{FOM}{figures of merit}
\newacronym{FIM}{FIM}{Fisher information matrix}

\newacronym[longplural={power spectral densities}]{PSD}{PSD}{power spectral density}
\newacronym[longplural={amplitude spectral densities}]{ASD}{ASD}{amplitude spectral density}
\newacronym[longplural={cross spectral densities}]{CSD}{CSD}{cross spectral density}
\newacronym{RMS}{RMS}{root mean square}
\newacronym{SNR}{SNR}{signal-to-noise ratio}
\newacronym[longplural={discrete Fourier transforms}]{DFT}{DFT}{discrete Fourier transform}
\newacronym{MCMC}{MCMC}{Markov chain Monte Carlo}
\newacronym{PAAM}{PAAM}{point-ahead angle mechanism}
\newacronym{PTMCMC}{PTMCMC}{parallel tempering Markov chain Monte Carlo}

\begin{document}

\title{{Extracting gravitational wave signals from LISA data in the presence of artifacts}}

\author{Castelli, Eleonora$^{1,2,3}$,
Baghi, Quentin$^4$, 
Baker, John G.$^2$,
Slutsky, Jacob$^2$,
Bobin, Jérôme$^5$,
Karnesis, Nikolaos$^6$,
Petiteau, Antoine$^5$,
Sauter, Orion$^7$,
Wass, Peter$^7$,
Weber, William J.$^8$}

\address{$^1$Center for Space Sciences and Technology, University of Maryland, Baltimore County, Baltimore, MD 21250, USA}
\address{$^2$Gravitational Astrophysics Lab, NASA Goddard Space Flight Center, Greenbelt, MD 20771, USA}
\address{$^3$Center for Research and Exploration in Space Science and Technology, NASA/GSFC, Greenbelt, MD 20771, USA}
\address{$^4$Astroparticule et Cosmologie, Université Paris Cité, 75205 Paris Cedex 13, France}
\address{$^5$IRFU, CEA, Universite Paris-Saclay, F-91191, Gif-sur-Yvette, France}
\address{$^6$School of Physics, Aristotle University Thessaloniki, 54006 Thessaloniki, Greece}
\address{$^7$Department of Mechanical and Aerospace Engineering, University of Florida, Gainesville, FL 32611, USA}
\address{$^8$Dipartimento di Fisica, Università di Trento and Trento Institute for Fundamental Physics and Application/INFN, 38123 Povo, Trento, Italy}

\date{\today}

\begin{abstract}
The Laser Interferometer Space Antenna (LISA) mission is being developed by ESA with NASA participation.  As it has recently passed the Mission Adoption milestone, models of the instruments and noise performance are becoming more detailed, and likewise prototype data analyses must as well. Assumptions such as Gaussianity, stationarity, and data continuity are unrealistic, and must be replaced with physically motivated data simulations, and data analysis methods adapted to accommodate such likely imperfections. To this end, the LISA Data Challenges have produced datasets featuring time-varying and unequal constellation armlength, and measurement artifacts including data interruptions and instrumental transients. In this work, we assess the impact of these data artifacts on the inference of Galactic Binary and Massive Black Hole properties. 
{Our analysis shows that the treatment of noise transients and gaps is necessary for effective parameter estimation, as they substantially corrupt the analysis if unmitigated. We find that straightforward mitigation techniques can significantly if imperfectly suppress artifacts. For the Galactic Binaries, mitigation of glitches was essentially total, while mitigations of the data gaps increased parameter uncertainty by approximately 10\%. For the Massive Black Hole binaries the particularly pernicious glitches resulted in a 30\% uncertainty increase after mitigations, while the data gaps can increase parameter uncertainty by up to several times. Critically, this underlines the importance of early detection of transient gravitational waves to ensure they are protected from planned data interruptions. }

\end{abstract}


\maketitle

\section{\label{sec:intro}Introduction}

By measuring \glspl{GW} in milliHertz frequencies \gls{LISA} will initiate a revolution comparable to the advent of infrared astronomy in the middle of the nineteenth century. Scheduled for launch in the mid-2030s, it will be sensitive to wavelengths ten thousand times larger than what we can currently observe from the ground. The LISA \gls{GW} sources will mainly be binaries of compact stars in our Galaxy, emitting quasi-monochromatic \glspl{GW}. Additionally, there is a high probability to observe transient gravitational radiation coming from the merger of distant \glspl{MBHB} with masses between $10^4$ and $10^7$ solar masses~\cite{colpi2024lisa}, \glspl{EMRI}, and stochastic backgrounds of \glspl{GW} from cosmological or astrophysical origins.

The \gls{ESA} adopted the mission in January 2024, kicking off the full design and implementation of the instrument and spacecraft. To prepare the mission adoption review, many studies were conducted regarding instrumentation, calibration, noise reduction, and data analysis. Among these, one team was responsible for assessing the robustness of the data analysis in the presence of data artifacts. The aim was to evaluate whether the mission science objectives could still be met in the presence of instrumental disturbances in the measurement process. In this article, we report the main findings of this pre-adoption study.

The \gls{LISA} measurements will be affected by various instrumental perturbations on top of the Gaussian stationary noise characterized by the \gls{PSD} function derived from the performance model. Based on the results obtained by the \gls{LPF} precursory mission, we foresee that two major instrumental artifacts impacting LISA will be gaps and glitches. The former are either temporary corruptions or complete interruptions of the science data stream, as those occurring at various points in the \gls{LPF} data~\cite{PhysRevLett.120.061101,10.1093/mnras/stz1017}. The latter are spurious instrumental transients arising in the data, as the force events transferring impulse to the \glspl{TM} the case of \gls{LPF}~\cite{PhysRevD.106.062001}. Some disturbances can be anticipated during flight, such as planned interruptions for routine observatory maintenance, while others appear randomly in the measurement process. It is paramount to evaluate their impact on extracting scientific information, and to which extent they can be mitigated.

Several approaches have been proposed to limit the effect of artifacts in the data analysis. Considerable efforts have been made to detect, identify and classify glitches in terrestrial detectors, often based on machine learning (see, for example, \cite{robinet_omicron_2020,Colgan2023,Alvarez-Lopez_2024}). Once detected, glitches can be modeled alongside the \gls{GW} signals to avoid bias in signal characterization~\cite{Bayeswave,Ashton2023}. This methodology has also been investigated to process future LISA measurements using wavelets~\cite{Robson2019}, parametric models and shapelets~\cite{spadaro2023glitch}, or machine learning~\cite{houba_detection_2024}. Concerning data gaps, special processing methods have also been devised in the context of ground-based detectors both in the case of masked disturbances~\cite{Zackay2021} and the absence of data~\cite{Dreissigacker2020}. However, they may be more challenging to deal with in LISA measurements where continuous sources are guaranteed and transient sources are observed over months. To mitigate their impact, data imputation techniques like Bayesian data augmentation~\cite{Baghi_2019}, sparse inpainting~\cite{blelly2021sparse}, or autoencoders~\cite{mao_novel_2024} have been prototyped.

The purpose of this work is not to introduce novel techniques to handle data artifacts. Rather, it firstly aims to assess the impact of instrumental disturbances on standard analysis techniques, with minor modifications to mitigate them. In this perspective, we address the \emph{Spritz} \gls{LDC}. As discussed above and in our conclusions, more advanced techniques have been proposed to deal with instrumental artifacts, and need to be further developed and incorporated in LISA global fit. Here, we report mainly on results that have been derived before LISA was adopted.

This study is based on simulated datasets released in the context of the publicly available \glspl{LDC}~\footnote{\url{https://lisa-ldc.lal.in2p3.fr/}}, a set of simulations aiming at testing data analysis pipelines to extract \gls{GW} with LISA~\cite{baghi2022}. The \gls{LDC}-2b challenge, {released in 2021 and} nicknamed \emph{Spritz}, currently is the only one that features noise artifacts in the form of gaps and glitches~\cite{spritz}. It is split in two main sub-challenges based on the included \gls{GW} source: the first one includes \glspl{VGB}, compact binary star systems that have been identified by electromagnetic observations to be potential \gls{LISA} source candidates (although this is under debate, see \cite{littenberg2024}). The second sub-challenge includes \glspl{MBHB}: one dataset contains a \gls{MBHB} with high \gls{SNR}, while the other dataset contains a lower-\gls{SNR} \gls{MBHB}. In this study, we restrict to the \gls{VGB} dataset and the loud \gls{MBHB} dataset. Each of the challenge datasets contain a predefined gap pattern and glitches modeled on to  \gls{LPF} observations.

\Sref{sec:data} describes the dataset content in terms of \gls{GW} sources and instrumental artifacts. In \Sref{sec:glitches}, we describe our approach to detect glitches and mitigate their overlap with \glspl{GW} through masking. In \Sref{sec:gaps}, we present our strategy to limit the impact of data gaps and masked glitches in the analysis. Then, we apply our method to the characterization of \glspl{VGB} in \Sref{sec:vgb} and of \glspl{MBHB} in \Sref{sec:mbhb}. \Sref{sec:conclusions} summarizes the main results of this study and discusses the future work that needs to be accomplished to increase the robustness of LISA data analysis methods against possible realistic instrumental conditions.

\section{\label{sec:data}Data features and artifacts}

{The two datasets under consideration are distributed as second generation \gls{TDI} Michelson combinations $X, Y, Z$ in the time-domain \cite{PhysRevD.59.102003, Armstrong_1999, PhysRevD.68.061303, PhysRevD.69.082001, PhysRevD.104.023006}. They }share the same instrumental noise simulation~\cite{bayle_lisanode}, Keplerian orbits simulation~\cite{bayle_orbits} and sampling time $\tau_s = \SI{5}{\second}$. They differ by their primary astrophysical source, duration, and injected glitches. Both publicly available \gls{LDC}-2b \emph{Spritz} datasets include the same simple periodic gap pattern of {7 hour} planned measurement dropouts {every 10 to 15 days} due to the periodic repointing of the high-gain communication antenna{, based on the information available at the time of the release}. In this study, we replace it with a more realistic scenario, {informed by the  up-to-date knowledge on the mission duty cycle available during the mission adoption review,} to investigate differences in gap occurrences and lengths.

The \gls{VGB} dataset contains a one-year observation of 36 VGB sources, corrupted by a realistic gap scenario and a population of glitches distributed according to the {\gls{LPF} glitch population analysis carried out in}~\cite{PhysRevD.105.042002}. 

\begin{table}[htbp]
    \caption{Subset of intrinsic \gls{MBHB} parameters used in the \gls{MBHB} dataset. Parameters are taken from the LDC-2b \emph{Spritz} documentation~\cite{spritz}: mass of the larger black hole $m_1$, mass of the smaller black hole $m_2$, the dimensionless spin of the larger black hole $a_1$, the dimensionless spin of the smaller black hole $a_2$, luminosity distance $D$, frequency at the innermost stable circular orbit $f_{\mathrm{ISCO}}$, total \gls{SNR}. The source stays in the LISA band for one month.}
        \lineup
    \label{tab:mbhb-parameters}
    \centering
    \rowcolors{0}{}{gray!20}
    \begin{tabular}{ll}
         \hline
         Parameter & Value  \\
         \hline
         $m_1$ [$10^6 M_\odot$] & \01.327 \\
         $m_2$ [$10^6 M_\odot$] & \00.612 \\
         $a_1$ & \00.7474 \\
         $a_2$ & \00.8388 \\
         $D$ [Gpc] & 13.47 \\
         $f_{\mathrm{ISCO}}$ [Hz] & $\00.01$ \\
         \hline
    \end{tabular}
\end{table}

The second dataset, called \gls{MBHB} dataset, features a one-month observation of a loud \gls{MBHB} signal (million-solar mass scale) with an \gls{SNR} of about 4370, corrupted by a realistic gap scenario and three short-duration loud glitches distributed in the inspiral, late inspiral and near merger parts of the signal. The \gls{MBHB} source parameters are listed in \Tref{tab:mbhb-parameters}.

We analyze each dataset separately and assess the impact of one class of artifacts at a time to determine their impact and possible mitigation techniques. 
For each noise artifact class, we present below the characteristics common to both datasets under analysis.

\subsection{Instrumental glitches}
Non-astrophysical transient signals were detected in the \gls{LPF} data~\cite{PhysRevLett.120.061101} and were referred to as \emph{glitches} during mission operations. They can originate from different locations inside the instrument and appear in multiple measurement channels~\cite{PhysRevD.106.062001}. The population analysis of LPF glitches~\cite{PhysRevD.105.042002} was used to simulate the class of events injected in the dataset under analysis, that is, force glitches acting on the \gls{LISA} \glspl{TM}.
\begin{figure}
    \centering
\includegraphics[width = 0.7\columnwidth]{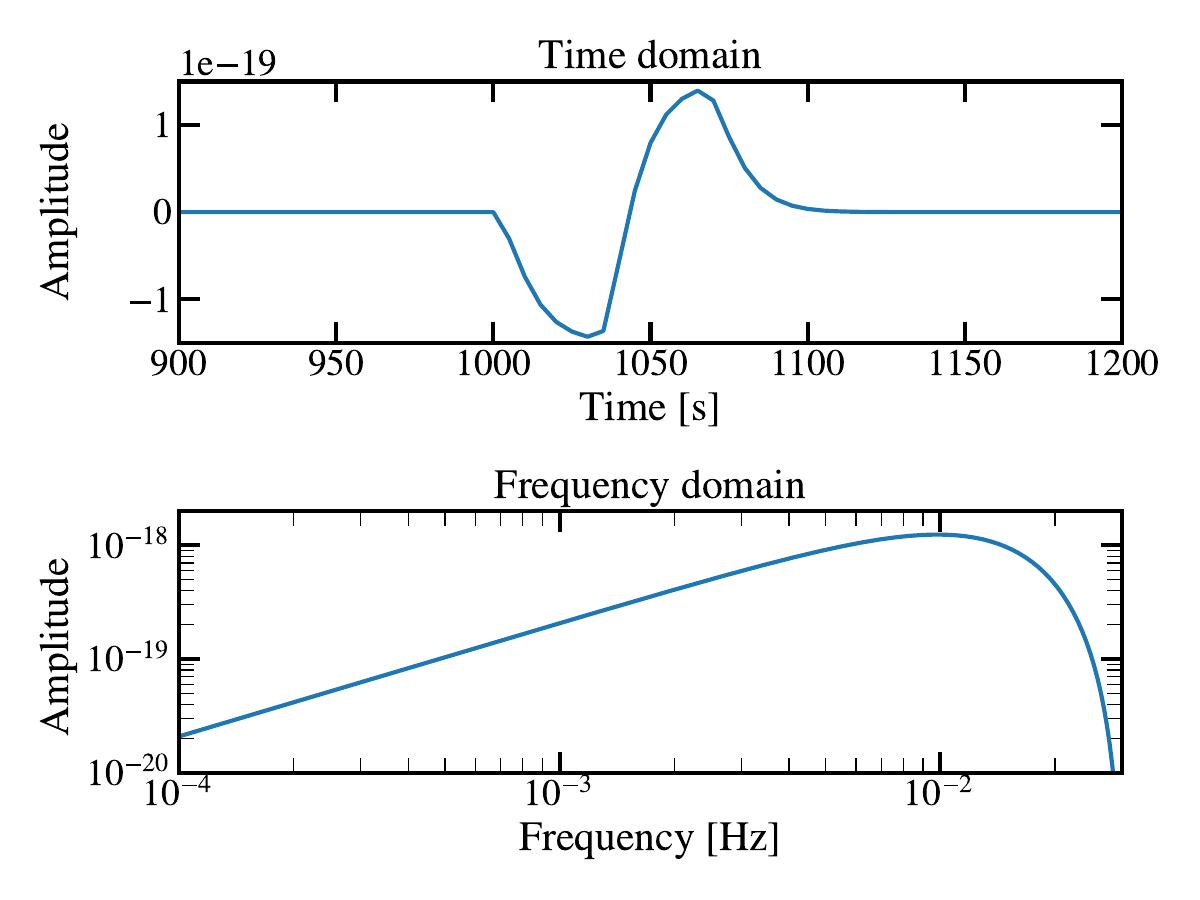}
    \caption{Example of how a force glitch acting on a \gls{TM} appears when measured through TDI. The time-domain TDI response to a force glitch can be modeled via \Eref{eq:phasemeter-glitch-2} with parameters corresponding to a damping time $\beta = 15$ seconds and an impulse $A = 4 \times 10^{-11} \mathrm{ms^{-1/2}}$.}
    \label{fig:glitch-example}
\end{figure}
In this scenario, the \gls{TDI} response to glitches exhibits a sharp up-down transient signal from the combination of delayed phasemeter measurements. These signals can carry a significant power relative to the Gaussian background noise. As shown in \Fref{fig:glitch-example}, glitches are very localized in time, their frequency response is spread across the band, with a peak of around \SI{10}{mHz}, and has a significant projection onto the \gls{GW} source signals if not correctly processed.

\begin{figure}
        \centering
        \includegraphics[width = 0.495\columnwidth]{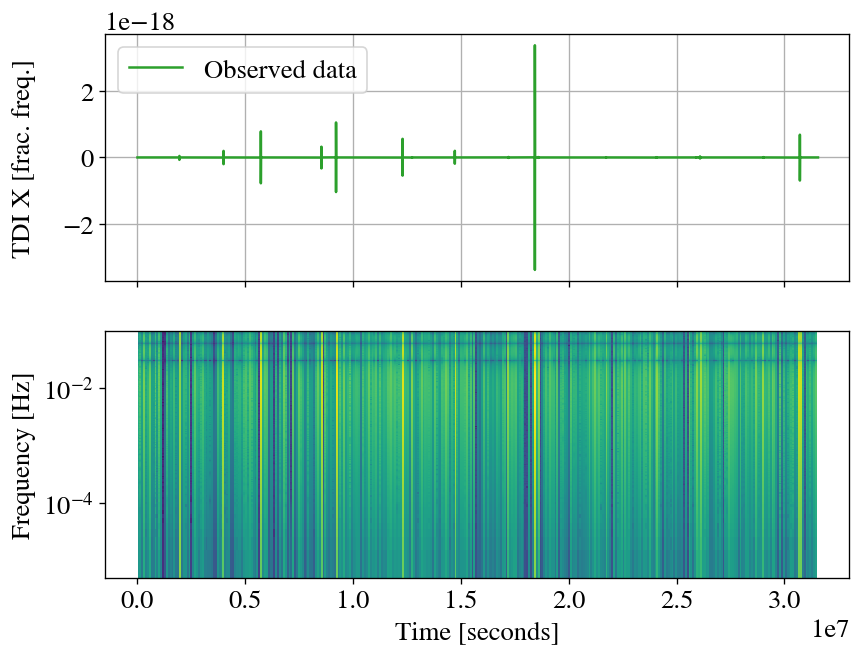}
        \includegraphics[width = 0.495\columnwidth]{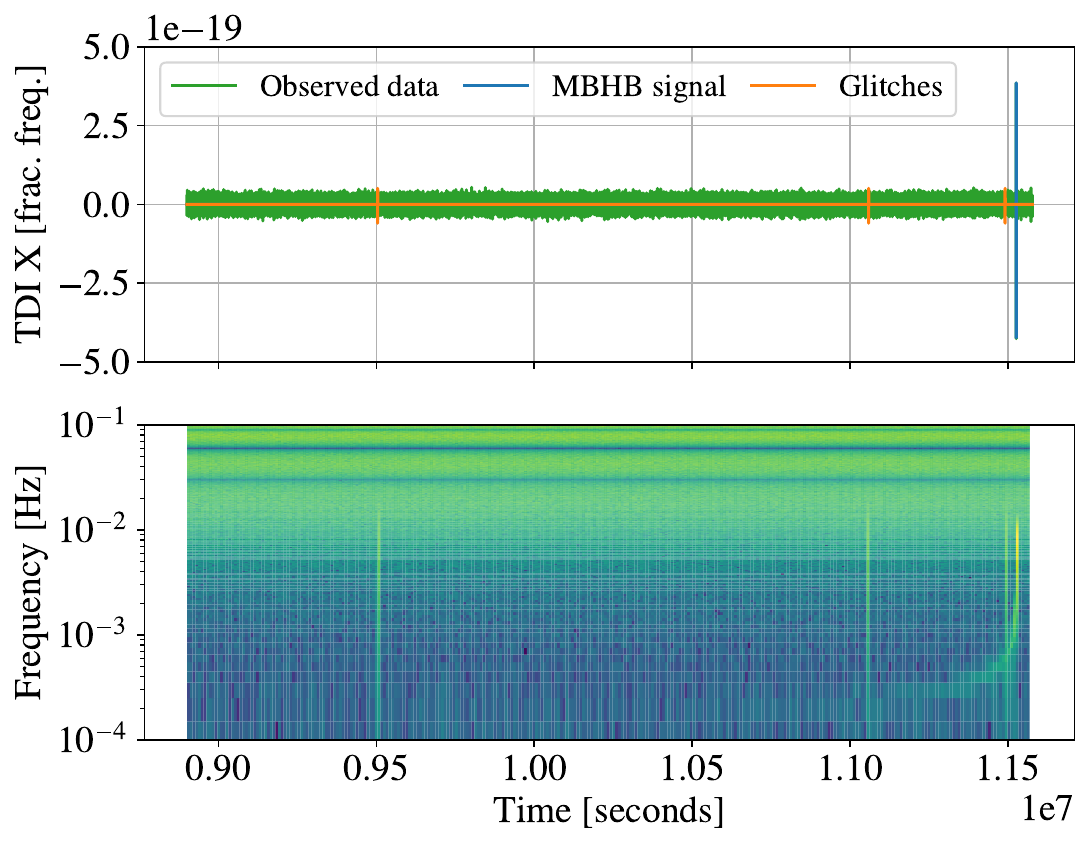}
        \caption{Top panels: \gls{TDI} $X$ time series of the \gls{VGB} (left) and \gls{MBHB} (right) datasets. For the \gls{VGB} dataset the observed data time-series containing noise, astrophysical signals and glitches is depicted in green, with only the bigger amplitude glitches visible in the plot. For the \gls{MBHB} dataset, we plot the time series of observed data containing noise, astrophysical signals and glitches (green),  the \gls{MBHB} signal (blue) and the injected glitches (orange). Bottom panels: spectrogram of the \gls{VGB}  (left) and \gls{MBHB} (right) datasets. The glitch occurrences are aligned with the top panel time axis. Notice the increased number of glitch occurrences (light vertical lines) visible in the \gls{VGB} spectrogram compared to the top panel.}
        \label{fig:TDI-time-series-glitches}
    \end{figure}

The time series and spectrogram of the VGB dataset containing the LPF glitch distribution are depicted on the left side of \Fref{fig:TDI-time-series-glitches}. The spikes in the top panel are the loudest glitches in the time-domain dataset (observable by eye), standing out of the noise. In the bottom panel, the spectrogram exhibits multiple vertical lines, most of which correspond to glitches we cannot distinguish in the top panel. The time series and spectrogram of the \gls{MBHB} dataset containing three identical glitch occurrences are depicted on the right side of \Fref{fig:TDI-time-series-glitches}. In that case, the three injected glitches appear as spikes in the upper and corresponding vertical lines in the bottom panels.



\subsection{Data gaps}
In the context of \gls{LISA}, a gap in the data stream is either due to missing telemetry or to a data portion highly contaminated by noise to the extent that it cannot be used in the data analysis. 


As mentioned above, in this study we discard the gap pattern of the publicly available datasets in favor of a more realistic realization of the gap pattern tied to the mission duty cycle, set near 82\%~\cite{colpi2024lisa}. The one implemented here is a plausible realization of a gap pattern including planned short-duration and unplanned long-duration data gaps.

Planned short-duration data gaps may occur when the measurements are too disturbed by maneuvers to be used for scientific purposes, e.g. during the periodic re-pointing of the high gain antenna towards Earth to establish communication and telemetry data, or the rotation of the \gls{PAAM} to correctly align the transmitting and receiving laser beams. Planned gaps will not be perfectly periodic, but will have~\emph{almost} constant cadence. Planned gaps may have moderate durations, such as are likely during antenna re-pointing gaps, lasting a few hours every 14 days, and the shorter duration \gls{PAAM} gaps lasting no more than 100 seconds with a rate of at most a few events per day~\cite{colpi2024lisa}.  The true set of planned interruptions will depend on final spacecraft and instrument design and operational details, but the pattern used in this analysis is designed to be plausible yet conservative.

Unplanned long-duration gaps are instead unpredictable interruptions in the data stream, caused by anomalies such as those observed in \gls{LPF}, e.g. recovery safe modes from spacecraft or instrument failures (such as computer crashes), and micrometeorite impacts~\cite{Thorpe_2019}. In our plausible yet conservative scenario, they can occur several dozens of times a year, and can last from one to three days.

\begin{figure}[htbp]
    \centering
    \includegraphics[width = 0.7\columnwidth]{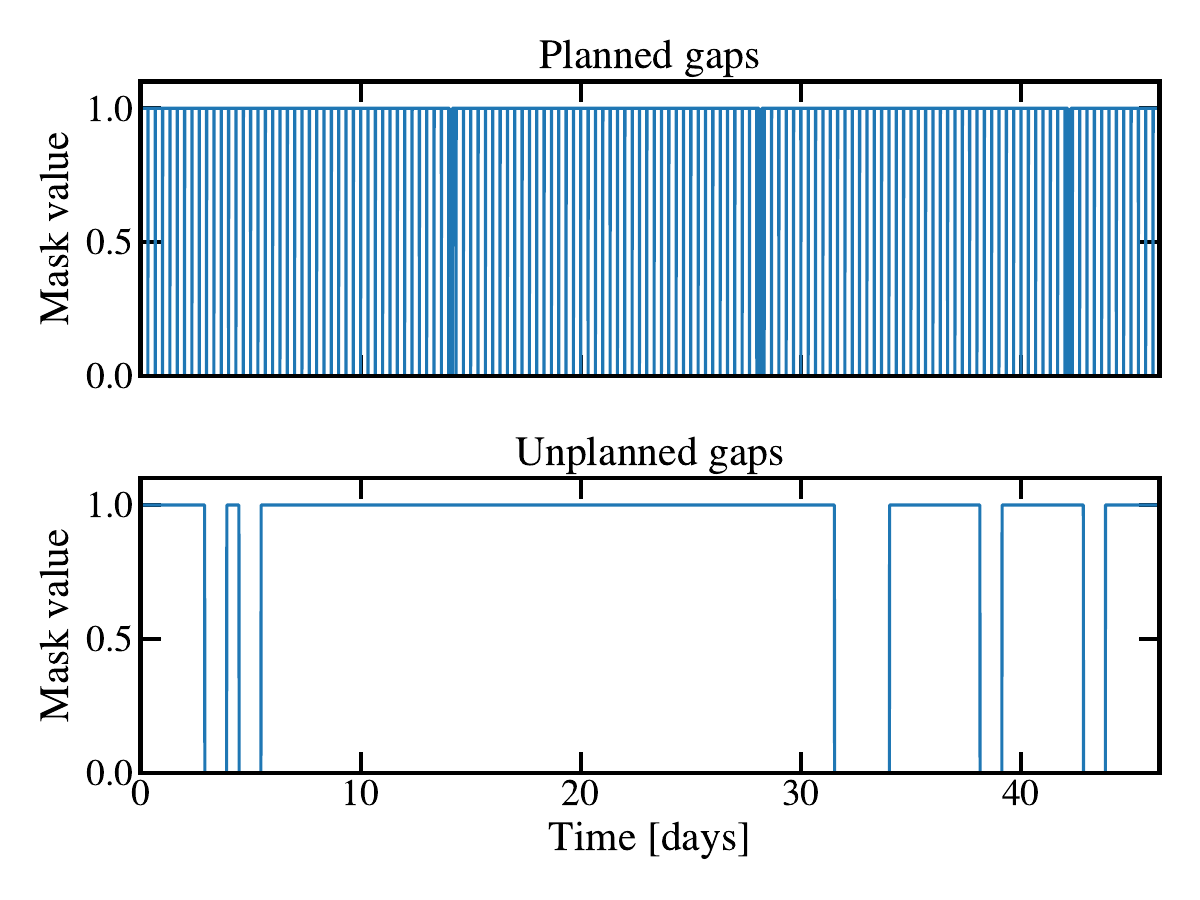}
    \caption{50-day extract from the year-long \gls{VGB} binary gap mask time series for planned (top panel) and unplanned (bottom panel) gaps. Planned gaps exhibit more frequent and shorter interruptions, while unplanned gaps are less frequent but longer.}
    \label{fig:gap-patterns}
\end{figure}

\begin{table}[htbp]
    \caption{Simulated gap occurrence frequencies and durations for planned (top) and unplanned (bottom) gaps.}
    \label{tab:gap-frequencies-durations}
    \rowcolors{0}{}{gray!20}
    \centering
    \begin{tabular}{lll}
         \hline
         Planned gap type & $\lambda_\text{per}$ [y$^{-1}$] & $\tau_\text{per}$ [h]\\
         \hline
        Short duration & 1095 & 0.028\\
        Medium duration & 26 & 3.3\\
         \hline
    \end{tabular}

    \vspace{0.2 cm}
    
        \begin{tabular}{lll}
         \hline
         Unplanned anomaly type & $\lambda_\text{anom}$ [y$^{-1}$] & $\tau_\text{anom}$ [h]\\
         \hline
         Platform interruptions & 3 & 60\\
         Payload interruptions & 4 & 66\\
         Environmental interruptions & 30 & 24 \\
         \hline
    \end{tabular}
\end{table}

In what follows, we simulate the planned and unplanned gap patterns by generating binary masks whose patterns are illustrated in \Fref{fig:gap-patterns}, and then applying them to non-gapped datasets.

The binary mask for planned gaps, in the top panel of \Fref{fig:gap-patterns}, takes into account the medium and short duration gaps mentioned above, and is generated by adding a slight randomization to the time in between two subsequent periodic gaps. {To simulate the \emph{quasi}-constant cadence of planned gaps, }the intra-gap period is defined as $1/\lambda_\text{per} \pm \tau_\text{rand}$, where $1/\lambda_\text{per}$ is the inverse of the periodic gap occurrence frequency, and $\tau_\text{rand}$ is a random fraction between 0 and 10\% of the gap duration $\tau_\text{per}$. The periodic gap occurrence frequencies {$\lambda_\text{per}$} and durations { $\tau_\text{per}$} for the planned gaps are listed in the top panel of \Tref{tab:gap-frequencies-durations}.

The binary mask for unplanned gaps, in the bottom panel of \Fref{fig:gap-patterns}, assumes data anomalies to be Poissonian uncorrelated, independent events occurring with a rate given by a characteristic yearly frequency $\lambda_\text{anom}$ for each kind of anomaly. Their arrival times differences follow an exponential distribution $\lambda_\text{anom} e^{-\lambda_\text{anom}}$, and their gap duration is $\tau_\text{anom}$.  The Poissonian gap occurrence frequencies and durations for the unplanned anomaly gaps are listed in the bottom panel of \Tref{tab:gap-frequencies-durations}.




\section{\label{sec:glitches}Glitch mitigation strategy}

In this section we discuss how glitches impact parameter estimation without mitigation. We then propose a way of detecting them, and we test a mitigation technique that consists of masking data spans affected by glitches, and treating them as gaps.

The strategy we adopt is threefold: i) detect glitches in the \gls{TDI} time series through matched filtering, ii) build a data mask from the detection results, iii) compute Fourier-transformed data by applying a smoothed version of the mask to mitigate leakage effects.

\subsection{Glitch detection}

To detect glitches in the data we use matched filtering, which requires an analytical model of the \gls{TDI} glitch response based on instrumental assumptions and expressed in the frequency domain, and a stationary noise PSD model. 

The time-domain glitch model assumed in the analysis is the one used for the injection in the \gls{LDC}-2b data \cite{spritz}, where the glitches are injected as independent Poissonian events with a rate of 4 events per day based on the population analysis ran in~\cite{PhysRevD.105.042002}. Each event results from a momentum perturbation on one of the six \glspl{TM}. The analytical TDI response to the time-domain glitch model is based on the propagation of a glitch signal on a single LISA link \cite{PhDMuratore}, extended to the case of TDI second generation {Michelson} combinations.

The noise \gls{PSD} model includes the stationary modeled instrumental noise sources for \gls{LISA} and the foreground of unresolved \gls{GB} sources that will be observable in the LISA band between \SI{0.2}{\milli\hertz} and \SI{5}{\milli\hertz}. The \gls{GB} foreground is modeled based on a population including all binaries with \gls{SNR} lower than 7 relative to the total noise budget~\cite{Karnesis2021}. The resulting stochastic process is non-stationary over time-scales longer than one month. The foreground non-stationarity is not an issue for the one-month \gls{MBHB} dataset, but needs to be taken into account in the case of the \gls{VGB} dataset: we estimate the noise PSD model for different month-long segments - a duration significantly shorter than the year-long full dataset. 

\begin{figure}[htbp]
    \centering
    \includegraphics[width = \columnwidth]{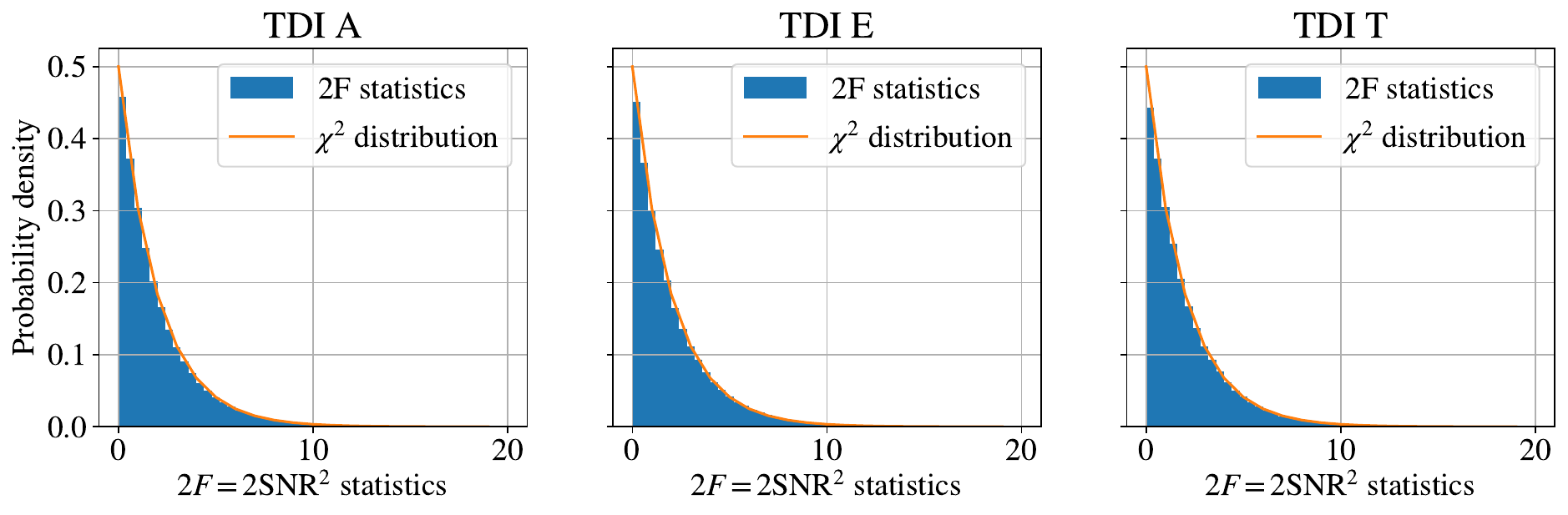}
    \caption{Distribution of $2\mathcal{F}$ for the one year VGB dataset containing only noise, for the econd generation pseudo-orthogonal $A, E, T$ TDI combinations (blue), compared to a $\chi^2_{2p}$-distribution with $2p$ degrees of freedom.}
    \label{fig:chi2-distribution}
\end{figure}

To implement matched filtering we {implement the same methods} of \cite{PhysRevD.105.042002}. While in the case of \gls{LPF} data only one channel was analyzed, here we extend the method to the multivariate detection problem involving the three {second generation pseudo-orthogonal} \gls{TDI} channels $A$, $E$ and $T$. Let us call $y_c$ the measured time series in channel $c$ with $c \in \{1, 2, 3\}$ corresponding to $A, E, T$, respectively. We use the $\mathcal{F}$-statistics as a detection quantity, 
\begin{equation}
\label{eq:multivariate-f-statistics}
\mathcal{F}(\boldsymbol{\theta}) = \sum_{c=1}^{p} \frac{1}{2} \frac{ \lvert \langle h_c | y_c\rangle \rvert^2}{\langle h_c | h_c \rangle},
\end{equation}
where $h_c$ is the glitch waveform in channel $c$. The waveform depends on parameter $\boldsymbol{\theta} = (A, t_0, \beta, I)$ including the glitch amplitude $A$, arrival time $t_0$, damping time $\beta$ and injection point $I$. 
We define the dot product as
\begin{equation}
\label{eq:dot-product}
    {\langle a_c | b_c \rangle} = \sum_{k={k_{\min}}}^{{k_{\max}}} \frac{\tilde{a}_c(k) \tilde{b}^{\ast}_c(k)}{S_c(f_k)},
\end{equation}
where $S_c$ is the one-sided noise \gls{PSD} in channel $c$, and $\tilde{a}$ denotes the Fourier transform of $a$. We define the estimated \gls{SNR} as the square root of the $\mathcal{F}$-statistics: $\mathrm{SNR} = \sqrt{\mathcal{F}}$. When considering $p$ different channels, the quantity $2 \mathcal{F}$ follows a $\chi^2_{2p}$ distribution with $2p$ degrees of freedom, {as shown in \Fref{fig:chi2-distribution} for the one year VGB dataset containing only noise, for the econd generation pseudo-
orthogonal $A, E, T$ TDI combinations.}

Following the same rationale as in \cite{PhysRevD.105.042002}, we can impose a false-alarm probability $\alpha$ by setting a threshold $F_0$ {on the $\mathcal{F}$-statistics $\mathcal{F} = F_0$} such that
\begin{equation}
\label{eq:false-alarm-threshold}
    P(p, F_0) =  (1 - \alpha)^{\frac{1}{q}} 
\end{equation}
where $q$ is the number of parameters we scan and $P(a, x)$ is the regularized lower incomplete gamma function.

To compute $h_c$, we write down a time-domain glitch model assuming a force perturbation on one of the test-masses, {integrated in time to be propagated through the phasemeter output in frequency as} 
\begin{equation}
\label{eq:phasemeter-glitch}
    v_g(t; A, \beta, t_0) = v_{g,0}(t - t_0 ; A, \beta),
\end{equation}
where
\begin{equation}
\label{eq:phasemeter-glitch-2}
    v_{g,0}(t; A, \beta) = \frac{2A}{\sqrt{\beta}} \left[ 1 - \left( 1 + \frac{t}{ \beta} \right) e^{\frac{-t}{ \beta}} \right] H(t).
\end{equation}
Here $H(t)$ is the Heaviside step function defined as $H(t) = 1$ for $t \geq 0$, $H(t) = 0$ otherwise, $A$ is the glitch amplitude, $\beta$ is the damping time and $t_0$ is the arrival time of the glitch.  The time-domain phasemeter model $v_{g}(t)$ follows the \texttt{Integrated Shapelet} model \footnote{Documentation available at \url{https://lisa-simulation.pages.in2p3.fr/glitch/v1.3/lpf.html}} implemented in the simulation tool \texttt{LISAGlitch (v1.0)} used to generate the glitch distribution in the LDC-2b datasets \cite{spritz, bayle_glitch}. 

{The injection point $I$ for the glitch is not explicitly defined within the glitch model: the phasemeter glitch model $v_g(t)$ is independent of the injection point, and it is added to the LISA single link identified by the target injection point I. That means that, } for a phasemeter glitch $v_g(t)$ injected on a specific \gls{TM} $I = \ $\texttt{tm\_12}{\footnote{the \gls{TM} located on spacecraft 1 and pointing at spacecraft 2, according to the notation in \cite{PhysRevD.107.083019}}}, the propagation of the glitch single link response {presented in \cite{PhDMuratore} can be evaluated for} the {second generation Michelson} \gls{TDI} variables $X, Y, Z$ {and result in the following glitch TDI waveform}
\begin{equation}
\label{eq:glitch-tdi-response}
\begin{split}
\eqalign
h_X(t) &= v_g(t) - 2 v_g(t - 4 T) + v_g (t - 8 T) \; ; \\
h_Y(t) &= - 2 v_g(t - T) + 2 v_g(t - 3 T)  \\
& \qquad + 2 v_g(t - 5 T) - 2 v_g(t - 7 T)\; ; \\
h_Z(t) &=0,
\end{split}
\end{equation}
where $T=8.3$ s is the average light travel time between the \gls{LISA} spacecraft. Note that the above model assumes that the transients are sufficiently short to consider constant and equal arms in the constellation, which is not true in the simulation used to generate the dataset. For glitches injected on different \glspl{TM}, {the usual cyclic permutations of the indices 1 → 2 → 3 → 1 apply~\cite{PhysRevD.104.023006}}. We then derive the {second generation} pseudo-orthogonal {TDI} variables $A, E, T$ from the {second generation} 
Michelson variables as 
\begin{equation}
   \begin{split} 
    A &= (Z - X)/\sqrt{2},\\
    E &= (X - 2Y + Z)/\sqrt{6},\\
    T &= (X + Y + Z)/\sqrt{3}.
    \end{split}
\end{equation}
{In what follows, TDI channels, variables or combination will be used to refer to the second generation pseudo-orthogonal TDI combinations.}

To allow for better computational efficiency, we analytically compute the \gls{DFT} of $h_c$ as a function of frequency (see~\ref{app:dfts}) to obtain the dot product in \Eref{eq:dot-product}.



\subsection{Application to LDC data}


{We first apply the \gls{SNR}-based matched filtering glitch detection method to a dataset containing only instrumental noise, galactic foreground and glitches, with no VGB signals, in order to test the algorithm's performance. Once confirmed that the method works as expected, we then apply it to the full \gls{LDC}-2b VGB observation dataset, which includes the astrophysical signals.}

{The method is applied identically to both the source-less dataset and to the full dataset. The results obtained from the full dataset analysis are used as input for the following steps of the analysis.} 
We divide the one-year time-domain dataset into 12 equal-length segments to reduce the computational cost. {The choice of splitting the one year dataset in 12 segments rather than 10 or 15 is not overly significant, and motivated by empirical testing of the algorithm performance: a significant increase in performance was not observed with increasing the number of segments.} Each month-long segment is treated independently, while the \gls{TDI} combinations $A$, $E$, and $T$ are analyzed jointly.

For each monthly segment, we estimate the noise \gls{PSD} via cubic spline interpolation of clean noise (glitch-less) stretches of data on a grid of minimally correlated frequencies. We average the \gls{PSD} estimation on all clean noise stretches longer than 7 days that start or end in the monthly segment. {This choice is a trade-off between it being long enough to reach low frequencies for the noise \gls{PSD} estimation, while at the same time allowing for averaging the PSD on at least 2 clean noise stretches within a one month long segment.} We choose the clean noise data stretches to minimize the impact of glitches on the noise \gls{PSD} estimation. They are identified after low-pass filtering of the year-long dataset, as the noise data stretches in-between glitches detected by preliminary sigma-clipping. The filter's cutoff frequency $f_\mathrm{cutoff} = \SI{1}{mHz}$ allows the removal of the high-frequency components of the \gls{LISA} noise. We calculate the minimally correlated frequency grid according to the method detailed in \cite{lpf2, armano2024indepth}: we obtain a log-spaced grid of 17 frequencies between $f_\text{min} = \SI{8}{\micro Hz}$ and $f_\text{max} = \SI{29}{\milli Hz}$: $f_\text{min} = (\SI{7}{days})^{-1}$ is determined by the minimum length of the clean noise data stretches, and $f_\text{max}$ by the location of the first dip in the LISA noise spectrum.

\begin{figure}
    \centering
    \includegraphics[width = 0.7\columnwidth]{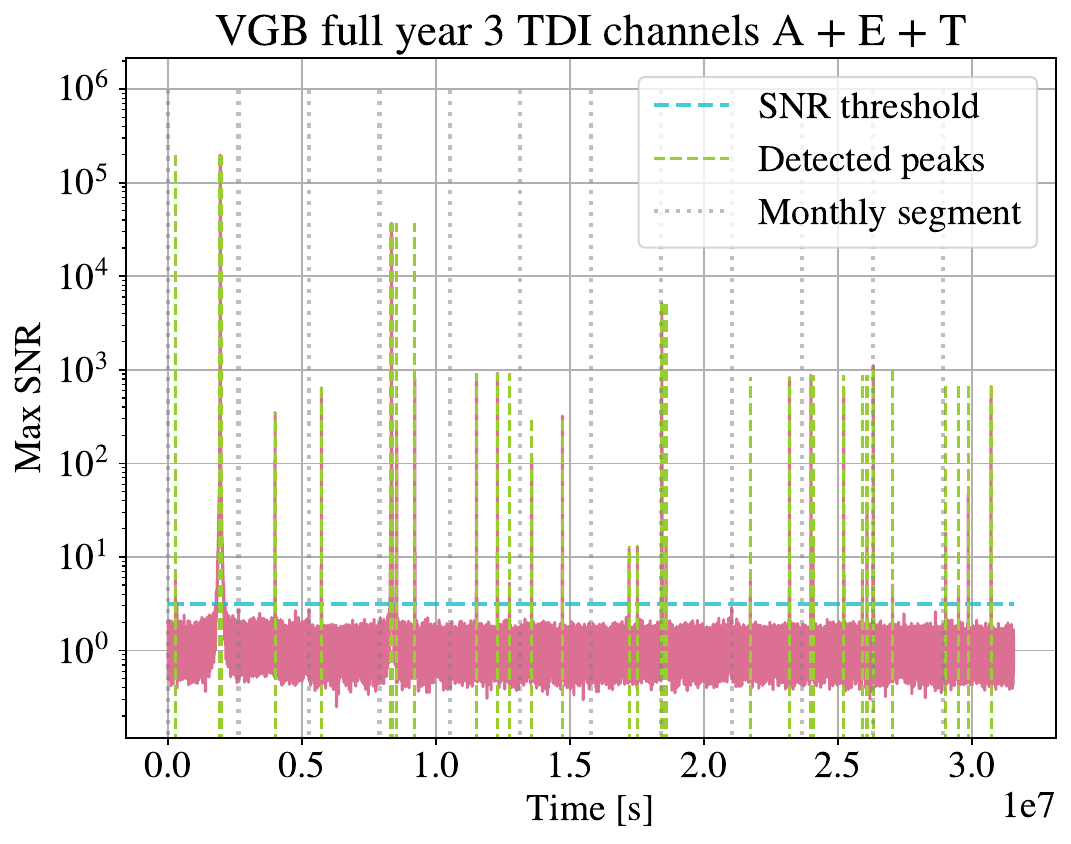}
    \caption{Maximum \gls{SNR} time series output by the \gls{SNR}-based matched filter applied to the  $A + E + T$ combination for the year-long \gls{VGB} dataset. The dataset is segmented in twelve monthly segments (separated by grey dotted vertical lines). The horizontal cyan dashed line is the {$\mathrm{SNR}_\text{th}^{AET}=\sqrt{F_0^{AET}}=2.74$} threshold corresponding to a $\alpha = 0.01$ false alarm probability. The vertical green dashed lines correspond to identified peaks in the maximum \gls{SNR}.}
    \label{fig:snr_aet}
\end{figure}

The \gls{SNR}-based matched filter is applied to the \gls{DFT} of each \gls{TDI} combination of the monthly segment data, with the PSD estimate as input and with a false alarm probability $\alpha = 0.01$, corresponding to {an SNR threshold $\mathrm{SNR}_\text{th} = \sqrt{F_0}=4.47$} in \Eref{eq:false-alarm-threshold}. We verify that $2 \mathcal{F}$ follows a $\chi^2$ distribution with $2p$ degrees of freedom for each monthly segment. We apply the \gls{SNR}-based matched filter to the combination of multiple \gls{TDI} representations too: $A + E$ and $A + E + T$, where the  $\alpha = 0.01$ false alarm probability corresponds to {$\mathrm{SNR}_\text{th}^{AE}=\sqrt{F_0^{AE}}=3.36$ and $\mathrm{SNR}_\text{th}^{AET}=\sqrt{F_0^{AET}}=2.74$} respectively. The maximum \gls{SNR} time series for $A + E + T$ is depicted in \Fref{fig:snr_aet}.

The detected peaks in the \gls{SNR} corresponding to glitches in the \gls{TDI} time series are then used as input for masking. 

The procedure could be applied analogously for the \gls{MBHB} dataset with a few precautions. First, lasting only one month, the dataset would not be divided into shorter segments. Second, the presence of a transient source could interfere with the \gls{SNR}-based matched filter, and a fraction of the \gls{MBHB}'s \gls{SNR} would be picked up by the matched filter: the development of statistical techniques to distinguish between glitches and \glspl{MBHB} is left to future studies. In what follows, the three glitch peaks in the \gls{MBHB} dataset were detected via sigma-clipping after low-passing data with a cutoff frequency $f_\text{c} = \SI{30}{mHz}$, which allows the removal of the higher frequency components of the \gls{LISA} noise. In our implementation of sigma-clipping for outlier detection, we evaluate the absolute deviation from the median of the low-passed data sample, and subsequently identifying peaks overcoming a certain threshold above that quantity. Glitches are detected as outliers if they rise above the median absolute deviation by a factor $5$, following the same \gls{SNR} detection threshold used in \cite{PhysRevD.105.042002}.

\subsection{Masking}

{After detecting the glitches in the full observation dataset in the previous section, }we proceed by masking, i.e. inserting gaps, in place of the detected glitches. Glitches are masked by zeroing out a specific-length data stretch around the identified glitch peaks. For the \gls{VGB} dataset, the width of each gap is set by the glitch duration parameter $\beta$ of  \Eref{eq:phasemeter-glitch-2} obtained from the matched filtering for each of the detected glitches, resulting in a binary mask with varying length. In the \gls{MBHB} dataset the three glitches are identical, and the width of each gap is set to be $n = 10$ data samples - corresponding to 50 seconds - before and after the glitch.

\begin{figure}
    \centering
    \includegraphics[width = 0.7\columnwidth]{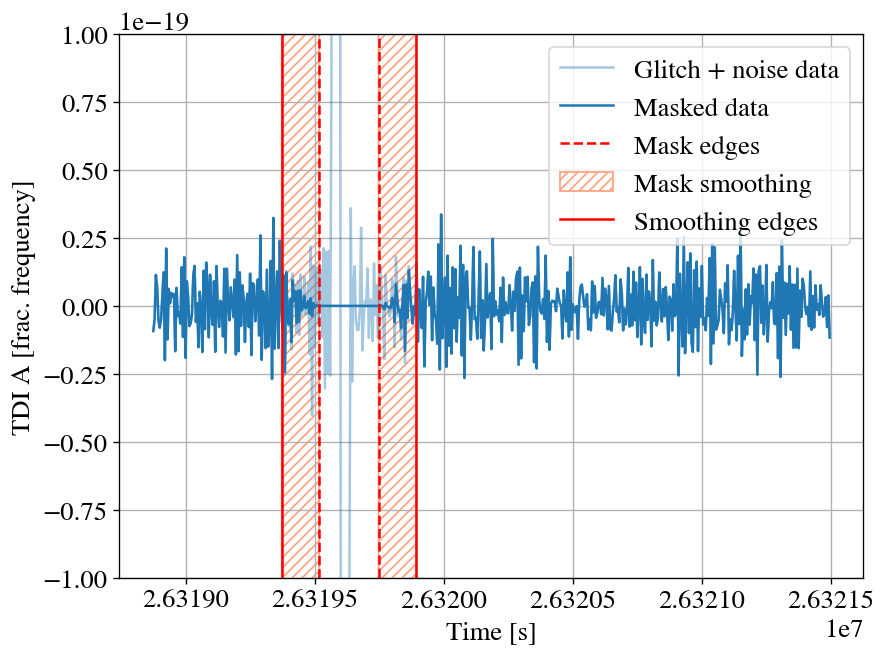}
    \caption{Example of glitch masking in the VGB dataset for a glitch occurring in the 10th monthly segment. Dark blue: TDI $A$ retained data with applied smoothed mask, identified as clean; light blue: data span identified as corrupted; red dashed lines: delimitation of the discarded data after masking; red {hatched} area: smoothing applied to the binary mask; red solid lines: delimitation of the smoothing area.}
    \label{fig:glitch-masking}
\end{figure}

After masking, the newly-gapped measurement data are treated following the gap mitigation strategy described in the next section. In the case of the VGB dataset under analysis where \gls{GW} sources are continuous, we apply smooth windowing to soften the mask edges between zero and one to mitigate the noise power leakage caused by the binary mask, as described in Subsection~\ref{sub:smooth-windowing}.
The masking and smoothing process is illustrated in \Fref{fig:glitch-masking} for a particular glitch event in the VGB dataset. The masked segment is plotted in light blue, the retained \gls{TDI} data in dark blue, the binary mask edges are marked by the red dashed lines, and the mask smoothing area is enclosed between the red solid and dashed lines and is represented by the red shaded vertical area.

\section{\label{sec:gaps}Gap mitigation strategy}

The LISA data analysis method we use operates in the frequency domain, where gaps have the main impact by causing noise frequency leakage inducing spectral distortions, as observed in \Fref{fig:noise-with-gaps-VGB}. If no mitigation is applied, these distortions severely impact the detection of continuous sources like \glspl{GB}. 
However, spectral leakage can be mitigated by applying smooth windowing in the time domain with continuous transitions at the gap edges before the Fourier transform. 
Smooth windowing allows for accurate parameter estimation for most of the \gls{GB} sources we consider, 
but it is nonetheless not applicable in the case of transient astrophysical sources like \glspl{MBHB}, because it distorts frequency-domain waveforms and can dramatically reduce \gls{SNR} due to smooth transitions.

\subsection{Smooth windowing}\label{sub:smooth-windowing}

In the case of continuous sources, the noise power leakage induced by gaps is mitigated by applying smooth windowing to the data, that is introducing progressive transitions at the binary mask edges between zero and one.

Let us call $w_n$ the smoothed mask time series. The windowed \gls{DFT} of any signal $h$, denoted as $\tilde{h}_w$, is computed as
\begin{eqnarray}
    \tilde{h}_w(f) = \tau_s \sum_{n=0}^{N-1} w_n h(t_n) e^{-2 i \pi f t_n} 
\end{eqnarray}
where $N$ is the time series size, $t_n = n \tau_s$ is the $n^{\mathrm{th}}$ time stamp, and $\tau_s$ is the sampling time. 

The effective \gls{SNR} of such a signal will be affected in two ways: i) loss of power due to missing data points and ii) noise power leakage from neighboring frequencies. Adopting the expression used in \gls{GW} data analysis, and following \cite{Baghi_2019}, we can write the effective \gls{SNR} as
\begin{equation}
\label{eq:effective-snr}
\mathrm{SNR}_{w} \equiv 4 \Delta f \int_{f_{\min}}^{f_{\max}} \frac{|\tilde{h}_{w}(f)|^2}{S_{\mathrm{eff}}(f)} df
\end{equation}
where $\Delta f = 1/(N\tau_s)$ is the frequency resolution and $S_\mathrm{eff}$ is the effective one-sided power spectral density, defined as
\begin{equation}
    S_{\mathrm{eff}}(f) = \int_{-f_s/2}^{+f_s/2} \left|\tilde{w}(f - f')\right|^2 S_n(f') df'
\end{equation}
 where $\tilde{w}$ is the Fourier transform of the window, $S_n$ is the one-sided \gls{PSD} of the noise, and $f_s = 1/\tau_s$ is the sampling frequency. This way, the numerator in \Eref{eq:effective-snr} is the windowed Fourier transform of the waveform, while the denominator is the noise \gls{PSD} convolved with the window's \gls{DFT}.

The mask smoothing is performed by choosing $w_n$ such that
\begin{equation}
\label{eq:binary-mask}
    w_n = \sum_{s=1}^{N_{\mathrm{seg}}} w_{N_s}(n-n_s),
\end{equation}
where $w_{N_s}$ designates the modified Hann window of length $N_s$~\cite{Carre2010}. The full window is therefore expressed as the sum of $N_{\mathrm{seg}}$ smooth windows of length $N_s$ starting at the time sample $n_s$ which are non-zero for samples between $n_s$ and $n_s + N_s$, and zero outside this interval. All time samples in between two consecutive segments (i.e., for which $n_s \leq n < n_{s+1}$) correspond to gaps. 

Explicitly, we have
\begin{equation}
\label{eq:gap-smoothing}
w_{N_s}(n) \equiv 
\begin{cases}
\frac{1}{2}\big[1-\cos \left( \frac{2 \pi n}{2 n_\text{wind}}\right)\big]  \quad\text{ if } 0 \leq n<n_\text{wind}; \\ 
1 \quad \text{ if } n_\text{wind} \leq n<N_s-n_\text{wind}; \\ 
\frac{1}{2}\big[1-\cos \left(\frac{2 \pi (n-N_s+2n_\text{wind})}{2 n_\text{wind}}\right)\big] \\ \qquad \qquad \text{ if } N_s-n_\text{wind} \leq n < N_s; \\ 
0 \quad \text { otherwise.}
\end{cases}
\end{equation}
We label $n_\text{wind}$ the transition length of the window, controlling the amount of smoothing we apply. We evaluate that $n_\text{wind} = 30$ samples (corresponding to a duration of 150 seconds) are an optimal trade-off between leakage mitigation and \gls{SNR} maximization. 

\begin{figure}
    \centering
    \includegraphics[width = 0.7\columnwidth, trim={2mm 2mm 2mm 0mm}, clip]{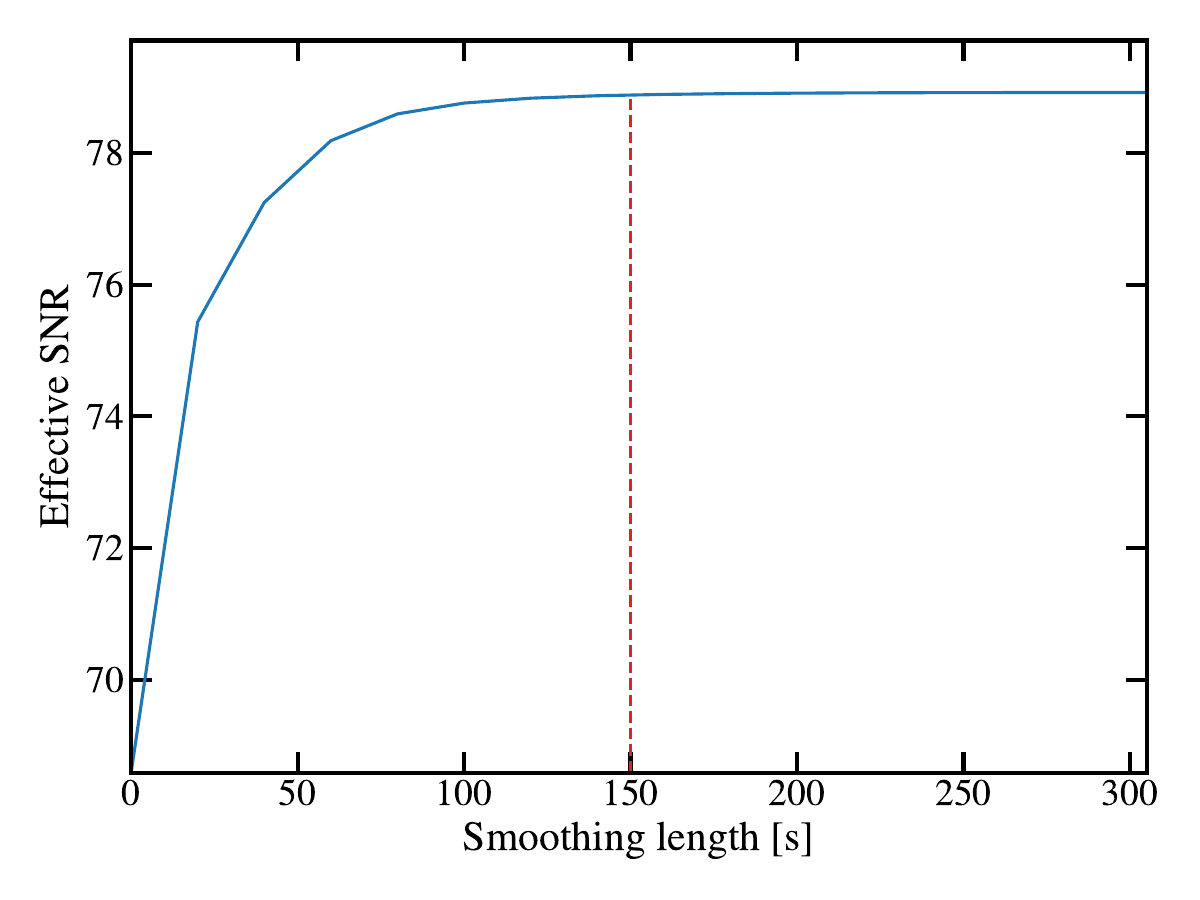}
    \caption{Effective \gls{SNR} of a VGB source at 1.8 mHz as a function of the gap smoothing length defined in \Eref{eq:gap-smoothing} and expressed in seconds. This curve illustrates the trade-off between leakage and mere loss of \gls{SNR}. At first, the \gls{SNR} improves with smoothing due to noise frequency leakage reduction. Then, it reaches a plateau where no more significant improvement is observed. The \gls{SNR} slowly decreases after some value because further smoothing the gaps leads to data loss. We choose to set the optimal smoothing equal to the value identified by the red dashed vertical line (150s).}
    \label{fig:smoothing-tradeoff}
\end{figure}

We illustrate this choice by plotting the effective \gls{SNR} computed with \Eref{eq:effective-snr} in \Fref{fig:smoothing-tradeoff}. The effective \gls{SNR} first increases with the transition length until it reaches a maximum before slowly decreasing. The regime where $S_{\mathrm{eff}}$ increases is dominated by noise effects: it corresponds to the distortion of the \gls{PSD} in the denominator of \Eref{eq:effective-snr} being mitigated by smoothing. Once enough smoothing is applied to mitigate leakage, the effective \gls{SNR} starts to decrease as the waveform in the numerator is smoothed out. The optimal smoothing length values {are} chosen around the transition between the two regimes. 

In most cases, smooth windowing allows for a sufficient reduction of leakage effect at a negligible cost in terms of \gls{SNR}, although it has been shown that depending on the noise \gls{PSD} and the source frequency, it may not be sufficient~\cite{Baghi_2019}. We assess this improvement in the following sections.

\section{\label{sec:vgb}Galactic binary characterization}

\subsection{Effect of glitches}
\subsubsection{Impact on noise spectrum}

We can {visually} evaluate the impact of glitches with the smooth masking mitigation technique by examining the Amplitude Spectral Density $\sqrt{\text{PSD}}$ of the masked \gls{TDI} time series. We compute the periodogram of the year-long VGB time series following Welch's periodogram implementation in the \texttt{scipy} Python library \cite{Welch1967,scipy}, choosing a Blackman averaging window of length $n_\text{win} = 256^2$ to suppress noise leakage.

\begin{figure}
    \centering
    \includegraphics[width = 0.8\columnwidth]{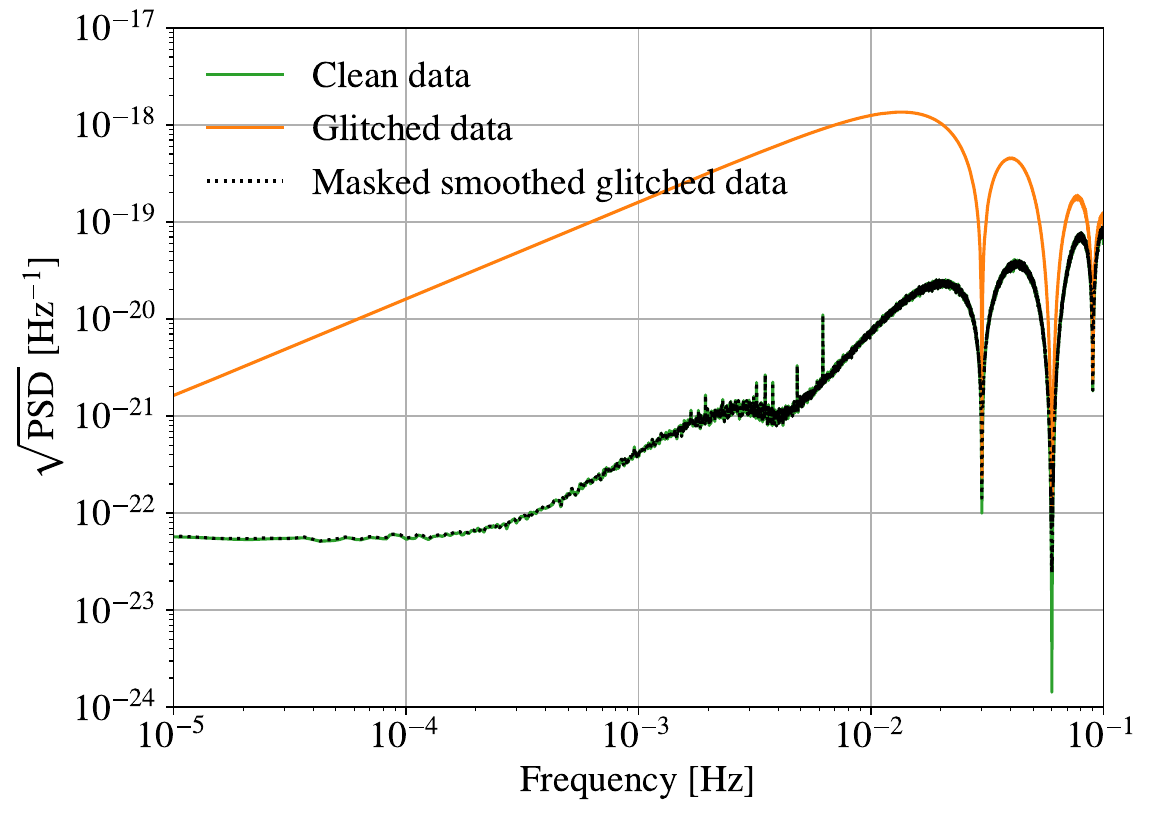}
    \caption{Comparison between the Amplitude Spectral Density $\sqrt{\text{PSD}}$ of data with (dashed black) and without (orange) glitch mitigation. The green spectrum corresponds to the clean data (baseline without glitches). All periodograms are computed for a one-year\gls{TDI} channel $A$ times series, with a {Blackman} averaging window length of $n_\text{win} = 256^2$.}
    \label{fig:noise-with-without-glitches-vgb}
\end{figure}

 In \Fref{fig:noise-with-without-glitches-vgb} we plot the $\sqrt{\text{PSD}}$ periodogram of \gls{TDI} channel $A$ as a function of frequency $f$ in the \gls{VGB} glitch scenario. 

The green spectrum corresponds to the baseline data without glitches, including only instrumental noise, galactic confusion noise and astrophysical \gls{VGB} signals identifiable as the spectral peaks between \SI{1}{mHz} and \SI{10}{mHz}. As a visual example of the glitches' contribution to the data stream, we compute the periodogram of the data including the population of LPF-like glitches, shown in orange, which rises one order of magnitude above the \gls{VGB} signals. Glitch events occur throughout the observation and their power piles up, overwhelming the signal. The comparison highlights the {necessity for accounting for glitches in the analysis}. 
The periodogram of the masked glitched data with applied smooth windowing is plotted as a dotted black curve. We observe a significant improvement compared to the raw glitched data (orange), with an effective power reduced to the level of the clean data (green).

\subsubsection{Impact on VGB parameter estimation}\label{sub:vgb-parameter-estimation-glitches}


We consider the 16 detectable sources out of the 36 \glspl{VGB} in the dataset, that is, with an \gls{SNR} larger than 7 for a one-year observation time relative to the instrumental noise and Galactic foreground. We perform a Bayesian inference of the sources parameters from clean and smoothly masked (mitigated) glitched data. The intrinsic source parameters are the frequency $f_0$, the frequency derivative $\dot{f}_0$, and sky location angles, including ecliptic longitude $\lambda$ and latitude $\beta$. The other parameters (amplitude $A$, inclination $i$ and polarization angles $\psi$, initial phase $\phi_0$) are parametrized as extrinsic amplitudes and will not be discussed. 

The inference is made using the stochastic sampling Python package \texttt{ptemcee}, which is a parallel tempering \gls{MCMC} based on affine-invariant ensemble sampling~\cite{Vousden2015,foreman-mackey_emcee_2013}. The posterior is explored by 32 chains, where annealing is applied through 20 different temperatures. The algorithm is run for $3  \times 10^5$ steps, and we keep the last 500 steps with a thinning of 6 out of 1 sample of each converged chain to derive the final parameter posteriors.



\begin{figure}
    \centering
    \includegraphics[width = 0.495\columnwidth]{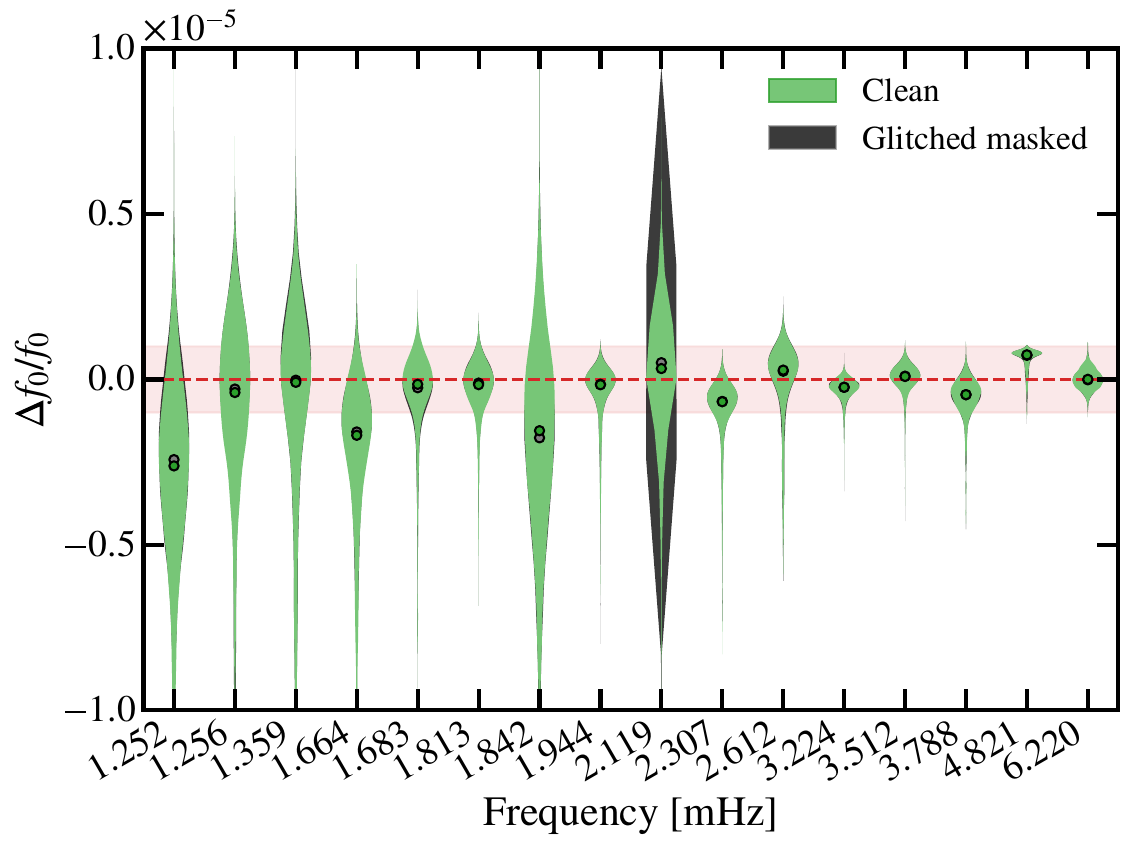}
        \includegraphics[width = 0.49\columnwidth]{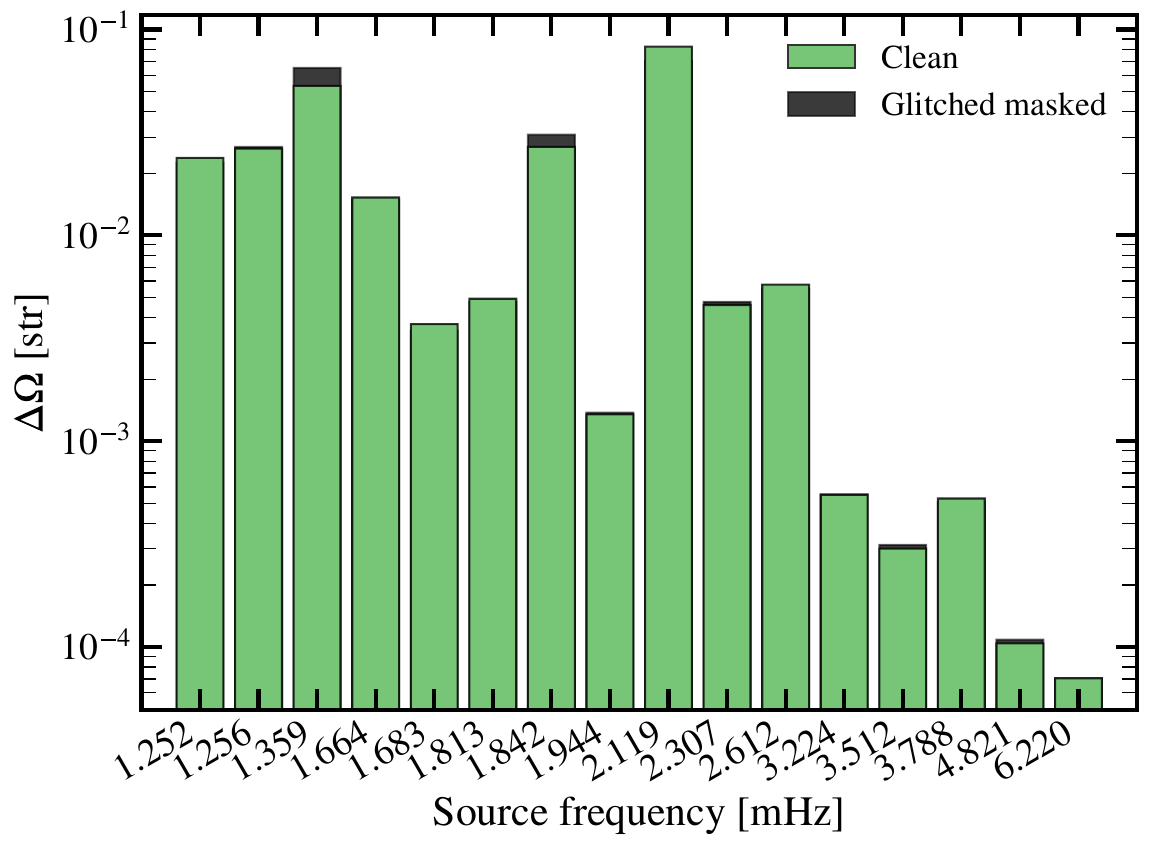}
    \caption{Left: Marginalized posterior distributions of \gls{VGB} frequencies expressed as errors relative to the injected value, obtained from clean data (green) and masked, glitched data (black) as a function of the source frequency. The colored dots indicate the medians of the distributions. The red dashed horizontal line indicates the injected value (zero error). The red shaded area is the region where $\Delta f_0 / f_0 < 10^{-6}$. Right: Solid angle error computed from sky location parameters posteriors obtained with clean data (green) and glitch masking (black).} 
    \label{fig:posterior-frequencies-locations}
\end{figure}

 We first consider the impact of glitches on the source frequencies, which are gathered in the left panel of \Fref{fig:posterior-frequencies-locations}.  The prior used for source frequency in the parameter estimation is a uniform prior with bandwidth of $\SI{1}{\milli\hertz}$ around the central frequency.  For most sources, the frequency posteriors obtained with the mitigated glitches (black) are very similar to the ones obtained with clean data (green). The masking technique is, therefore, efficient in mitigating the bias caused by glitches. The red area corresponds to relative frequency errors less than one part per million. This indication relates the result to the operational requirement on bright \glspl{GB}, which states that the relative error on the orbital period should be less than $10^{-6}$~\cite{ScRD_2018}. Hence, for this selection of sources, the number of sources verifying this requirement is not affected by glitches once they have been mitigated. 

Nevertheless, we observe a difference for the source at $f_0 = 2.11$ mHz where the posterior distribution appears broadened relative to the clean data result. This source has the lowest \gls{SNR} of all detectable \glspl{VGB}, close to the detection threshold ($\mathrm{SNR} = 7$). This result suggests that glitches may impact low-\gls{SNR} more than others, especially for frequencies with the strongest confusion foreground.

We also assess the impact on source sky localization by computing the solid angle error $\Delta \Omega$ defined as~\cite{cutler-98} 
\begin{eqnarray}
\label{eq:solid-angle-error}
   \Delta \Omega^2 = (2 \pi)^2\left(\operatorname{var}(\lambda) \operatorname{var}(\beta)-\operatorname{cov}(\lambda, \beta)^2\right),
\end{eqnarray}
where $\beta$ and $\lambda$ are the ecliptic latitude and longitude, respectively. We report the results in the right panel of  \Fref{fig:posterior-frequencies-locations}, where we plot values for $\Delta \Omega$ in green for the clean data set and grey for the masked, glitched dataset. Again, we obtain comparable results for clean and masked glitched data. The difference in sky location is insignificant, especially for the source at $f_0 = 2.119$ mHz where we observed a broader frequency posterior. {Note that in reality, \glspl{VGB}' sky locations will be known thanks to electromagnetic observations, but here we assume that they are estimated with \glspl{GW} alone.}

\subsection{Effect of gaps}

\subsubsection{Impact on noise spectrum}
\begin{figure}
    \centering
    \includegraphics[width = 0.8\columnwidth]{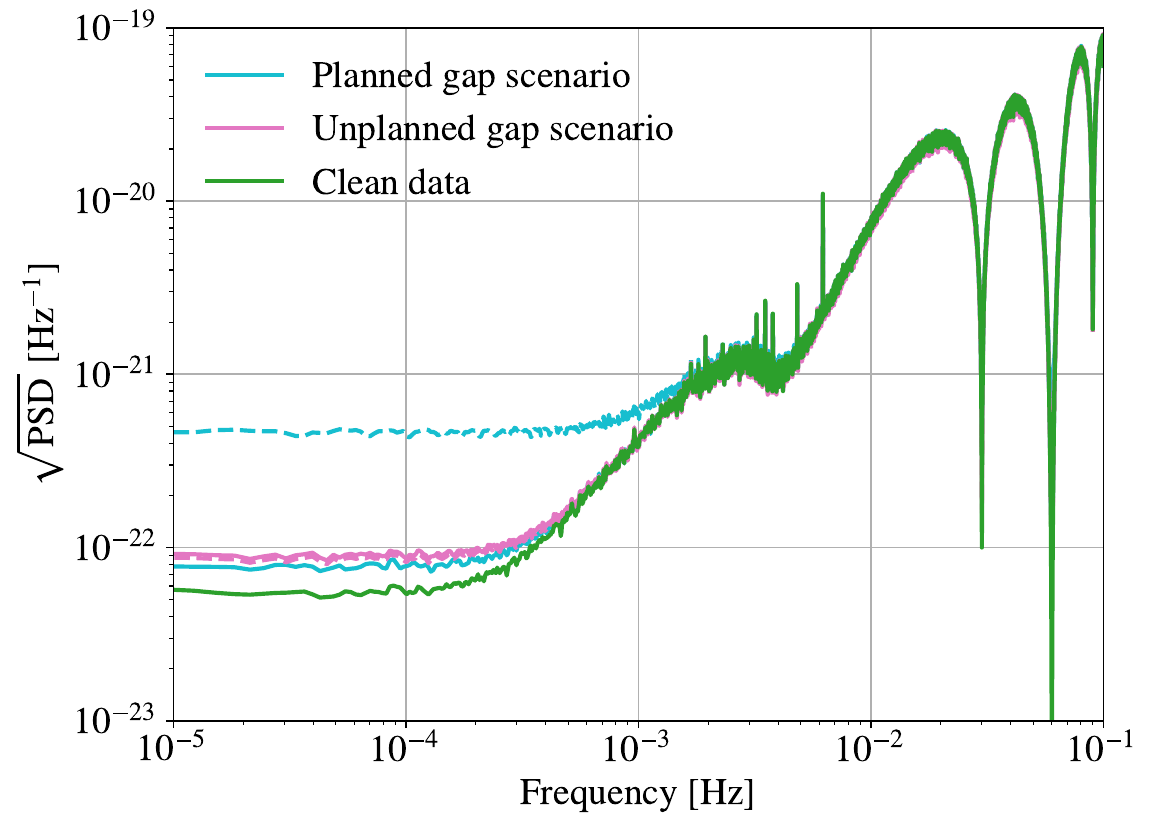} \ 
    \caption{Comparison between the Amplitude Spectral Density $\sqrt{\text{PSD}}$ of data with different gap patterns applied to the \gls{VGB} dataset: planned gap scenario (cyan) and unplanned gap scenario (pink). The green spectrum corresponds to the clean data (baseline without gaps). Dashed lines indicate data with no gap mitigation, solid lines indicate noise spectra with applied smooth windowing. All periodograms are computed for a one-year TDI channel $A$ time series, with an averaging window length of $n_\text{win} = 256^2$ and a Blackman averaging window.}
    \label{fig:noise-with-gaps-VGB}
\end{figure}

We now {visually} evaluate the impact of planned and unplanned gaps with the smooth masking mitigation technique by examining the periodograms of the \gls{TDI} time series.

In \Fref{fig:noise-with-gaps-VGB}, we plot the $\sqrt{\text{PSD}}$ periodogram of \gls{TDI} channel $A$ as a function of frequency $f$ in the \gls{VGB} glitch scenario. The green spectrum corresponds to baseline data without gaps, and includes Gaussian noise and {VGB} signals, which are identifiable by the spectral peaks visible between \SI{1}{mHz} and \SI{10}{mHz}. 

Introducing gaps in the time series and performing the same periodogram computation yields the planned gap scenario (cyan) and unplanned gap scenario (pink) spectra. For each gap scenario, we plot the periodogram of unmitigated data (dashed curves) and the periodogram of data after applying smooth windowing (solid curves). 

Both unmitigated gapped spectra plateau above the baseline spectrum below \SI{1}{mHz}. The unmitigated planned gap scenario (cyan) flattens one full order of magnitude above the clean data due to frequency leakage, whereas the unmitigated unplanned gaps spectrum is less noisy. This is mainly related to the short-duration planned gaps. Indeed, previous studies~\cite{Baghi_2019} show that short and frequent gaps generate more leakage than long and rare gaps. Leakage is dominant at low frequencies below 2 mHz due to the smearing effect of the  window's Fourier transform, acting as a convolution kernel on the noise \gls{PSD}. This convolution usually transfers power from higher-power regions to lower-power regions. The smoothed planned spectrum instead show a suppression of excess noise below \SI{1}{mHz}, and is comparable to the noise level of the smoothed unplanned gap scenario.

\subsubsection{\label{sec:vgb-mitigated-pe}Impact on parameter estimation}

We estimate all detectable source parameters in the presence of gaps, analogously to what we did for glitches in Sec.~\ref{sub:vgb-parameter-estimation-glitches}.

\begin{figure}
    \centering
            \includegraphics[width = 0.495\columnwidth]{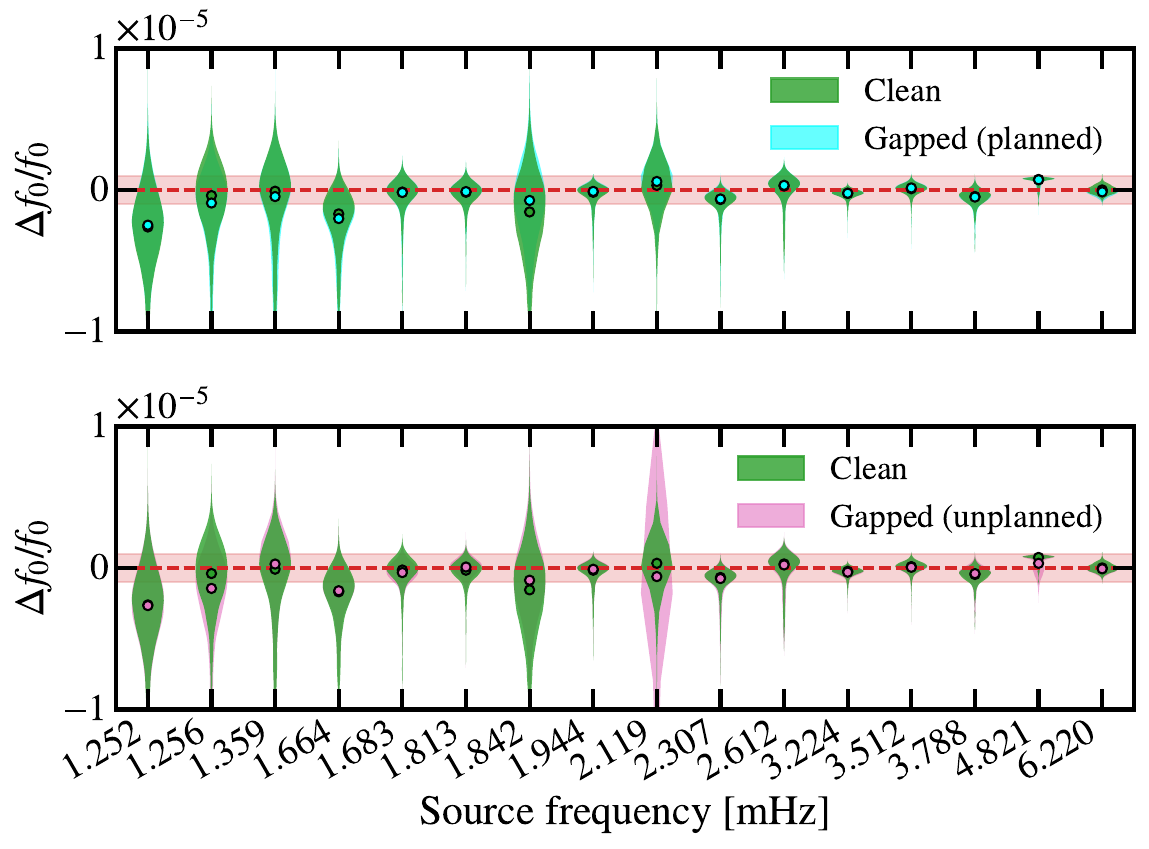} 
                \includegraphics[width = 0.495\columnwidth]{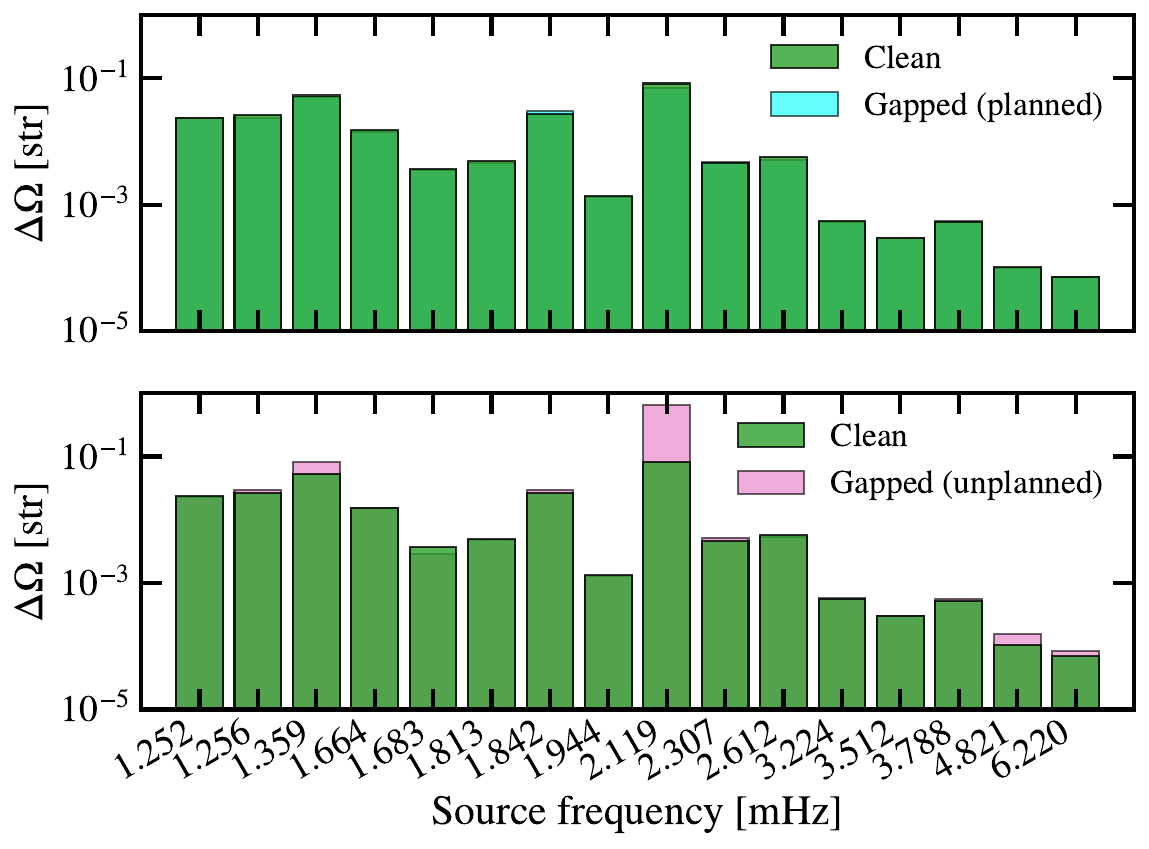}
    \caption{Left: Marginalized posterior probabilities for \gls{VGB} source frequencies using smooth windowing. Top left panel: Posteriors obtained with planned gaps (cyan), compared with clean data (green). Bottom left panel: Posteriors obtained with unplanned gaps (pink), compared with clean data (green). The colored dots indicate the medians of the distributions. The red dashed horizontal line indicates the injected value (zero error). The red shaded area is the region where $\Delta f_0 / f_0 < 10^{-6}$. Right: {VGB} sky location error (solid angle) using smooth windowing. Top right panel: Posteriors obtained with planned gaps (cyan), compared with clean data (green). Bottom right panel: Posteriors obtained with unplanned gaps (pink), compared with clean data (green).} 
    \label{fig:posterior-frequencies-gaps}
\end{figure}

We report the source frequency inference results in \Fref{fig:posterior-frequencies-gaps}. We don't observe a noticeable impact on any of the \gls{VGB} sources in the case of planned gaps (top panel). For unplanned gaps, we observe a more significant impact on the frequency posteriors, likely because this is the scenario with the largest data loss (about 10\%, with random occurrences of long gaps). The inference is mildly affected for most sources, where the uncertainty is increased by $10\%$, which is barely visible on the plot. The sources with frequencies $f_0 = 2.119$ mHz and $f_0 = 4.821$ mHz show the most significant impact. The posterior width of the former is significantly increased with the gapped data, while the posterior modes are distorted in the latter. 

We examine the sky location errors obtained with clean and gapped data in \Fref{fig:posterior-frequencies-gaps}, where we compute the errors in terms of solid angle as given by \Eref{eq:solid-angle-error}.
In the case of planned gaps (top panel), the sky location errors obtained with the smooth windowing method (blue bars) are comparable to the errors obtained from clean data (green bars). 

In the case of unplanned gaps (bottom panel), the sky location errors are usually slightly larger than the ones obtained with clean data. The largest degradation is observed for the source $f_0 = 2.119$ mHz. 

\section{\label{sec:mbhb}Massive black hole binary characterization}

\subsection{Effect of glitches}
\subsubsection{Impact on noise spectrum}

\begin{figure}
    \centering
    \includegraphics[width = 0.8\columnwidth]{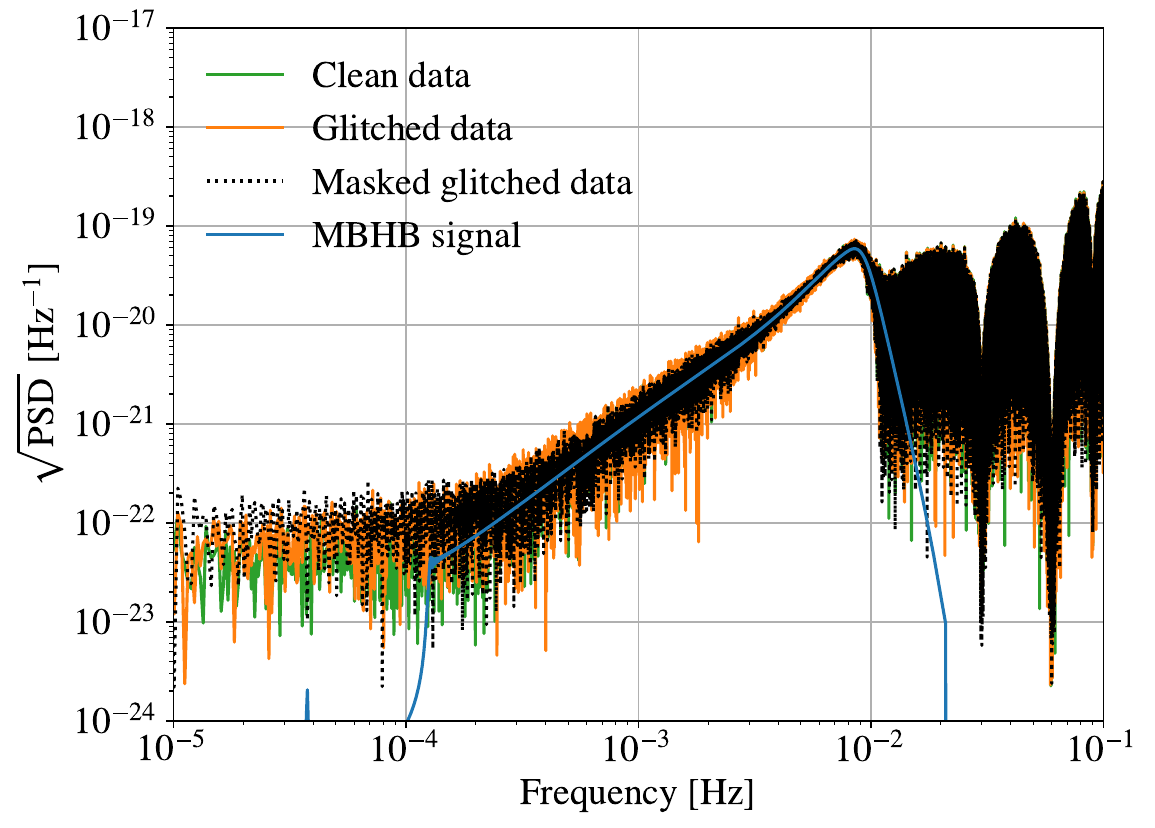} \\ 
    \caption{Comparison between the Amplitude Spectral Density $\sqrt{\text{PSD}}$ of data  with (dashed black) and without (orange) glitch mitigation for the \gls{MBHB} dataset. The green spectrum corresponds to the clean data (baseline without glitches). All periodograms are computed for a one-month \gls{TDI} channel $A$ time series, with no averaging.}
    \label{fig:noise-with-without-glitches-MBHB1}
\end{figure}

We now consider the \gls{MBHB} dataset.
In \Fref{fig:noise-with-without-glitches-MBHB1}, we perform the same comparison as in \Fref{fig:noise-with-without-glitches-vgb}: we plot the amplitude spectral density periodogram $\sqrt{\text{PSD}}$ of \gls{TDI} channel $A$ as a function of frequency $f$, with no averaging window. For transient sources we avoid tapering windows in the periodogram estimation because they significantly distort the signal spectrum and decrease the \gls{SNR}. 

In the \gls{MBHB} case, the improvement brought by the masking process is difficult to assess from the visual inspection of the spectra. The impact of glitches on the spectrum (orange) is mild compared to the clean data (green) and mostly affects frequencies below a few \si{mHz}. \Fref{fig:noise-with-without-glitches-MBHB1} suggests that some spurious power is reduced in the milliHertz band, at the cost of an increase of power at low frequencies due to noise leakage. Besides, the \gls{SNR} of the \gls{MBHB} signal (blue) is concentrated around the merger frequency ($\sim 8$ mHz), which is comparatively less impacted.

\subsubsection{\label{sec:glitch-mitigation-impact-on-mbhb-pe}Impact on parameter estimation}
\begin{figure}
    \centering
\includegraphics[width = 0.8\columnwidth]{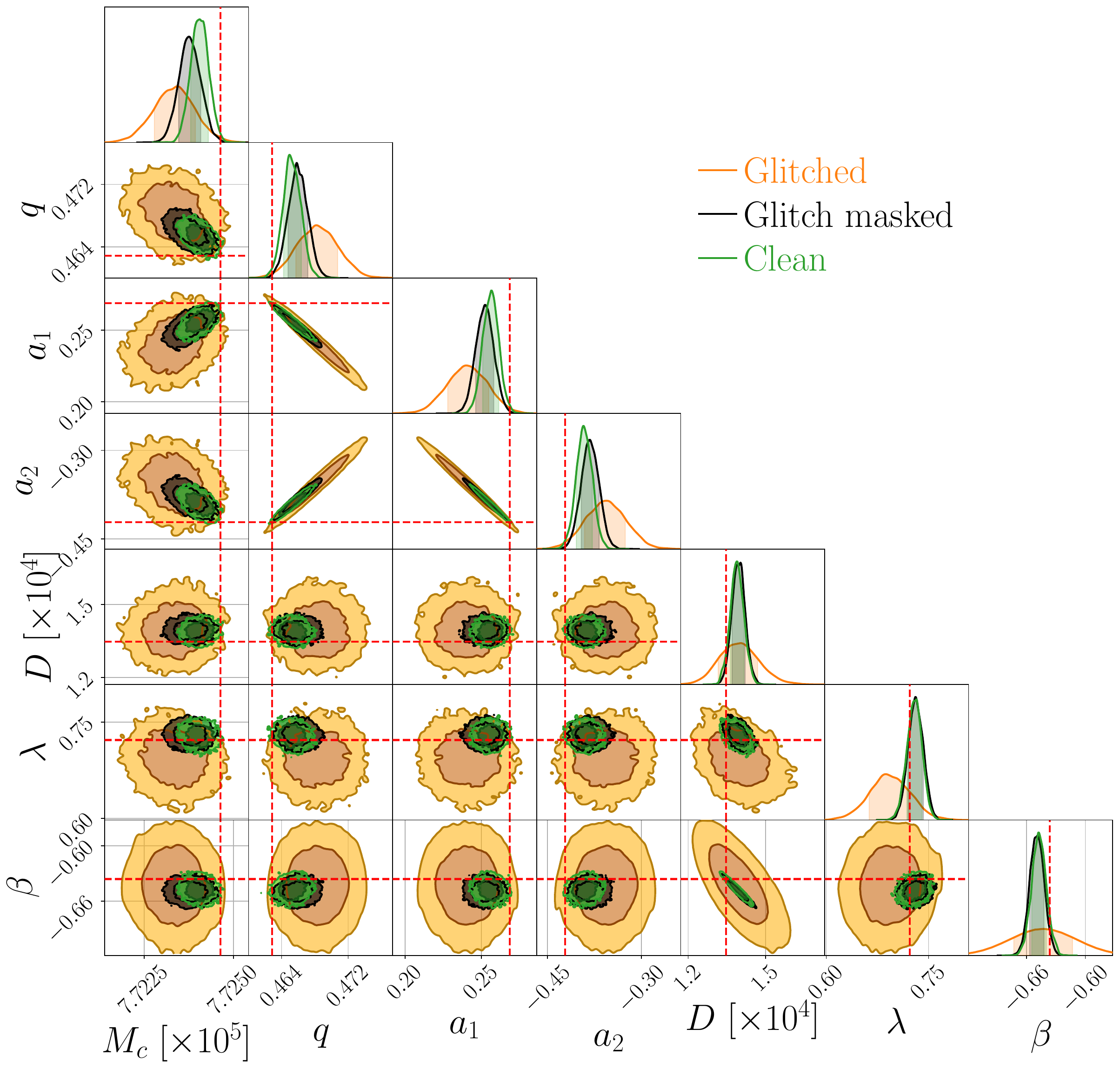}
    \caption{Corner plot of the \gls{MBHB} search parameters for the following datasets: clean data ({green}), glitch-masked data (black), and glitched data ({orange}). Depicted in the corner plots are the posterior probabilities with $1-\sigma$ and $2-\sigma$ confidence areas for the following parameters: chirp mass $M_c$, mass ratio $q$, dimensionless spin of the larger black hole $a_1$, dimensionless spin of the smaller black hole $a_2$, luminosity distance $D$, ecliptic longitude $\lambda$, ecliptic latitude $\beta$. 
    Red dotted lines mark the true parameter values: {the MCMC search was implemented with equal-armlength waveform approximation, while the \gls{LDC} data include an unequal-armlength LISA constellation \cite{spritz}, leading to a slight bias between the true parameter values and the clean data posterior.}}
    \label{fig:mbhb-posterior-glitches}
\end{figure}

We explore the posterior distribution of the \gls{MBHB} signal parameters using \texttt{lisabeta}\footnote{https://pypi.org/project/lisabeta/}   {and parallel tempering MCMC implemented in} \texttt{ptmcmc}\footnote{https://github.com/JohnGBaker/ptmcmc}~\cite{PhysRevD.103.083011}. The source is characterized by 11 parameters, including chirp mass $M_c$ [$M_\odot$], binary mass ratio $q$, spin parameters $a_1, \, a_2$, luminosity distance [\si{Mpc}], latitude $\beta$ and longitude $\lambda$ angles [\si{rad}] in the constellation frame. The dashed red lines show the injected parameters, also listed in \Tref{tab:mbhb-parameters}. We use \gls{TDI} channels $A$ and $E$ to perform the inference. {The inference analysis incorporated 32 parallel chains at temperatures tuned for consistent exchange. The calculation proceeded until the the number of cold chain steps (after a burn-in period) was 2000 times the longest estimated correlation length among all parameters.} 

We notice a slight bias of the posterior probability density with respect to the injection {marked by the red dashed lines in \Fref{fig:mbhb-posterior-glitches}} due to the high \gls{SNR} that makes it challenging to sample the posterior distribution. In addition, the \gls{MCMC} search was implemented using the equal-armlength waveform approximation while the \gls{LDC} data include an unequal-armlength LISA constellation~\cite{spritz}. The presence of this slight bias does not change the underlying meaning of the following discussion.

\Fref{fig:mbhb-posterior-glitches} shows a posterior probability corner plot for selected parameters, for each of the three \gls{MBHB} dataset scenarios: clean data ({green}), glitched data ({orange}) and glitch masked data (black). 
The inference with the raw, glitched data yields significantly wider parameter posteriors compared to the clean case. For example, the error on the chirp mass increases by a factor larger than 2, and we observe a slight shift compared to the clean (green) posterior. The same can be observed for sky location angles. Therefore, the presence of unmitigated glitches causes significant parameter precision and accuracy degradation while the source is fairly well recovered. 

However, gapping the three glitch occurrences in the \gls{MBHB} dataset allows us to mitigate this widening to about 30\%, as shown by the black posterior. 

\subsection{Effect of gaps}
\subsubsection{Impact on noise spectrum}
\begin{figure}
    \centering
    \includegraphics[width = 0.8\columnwidth]{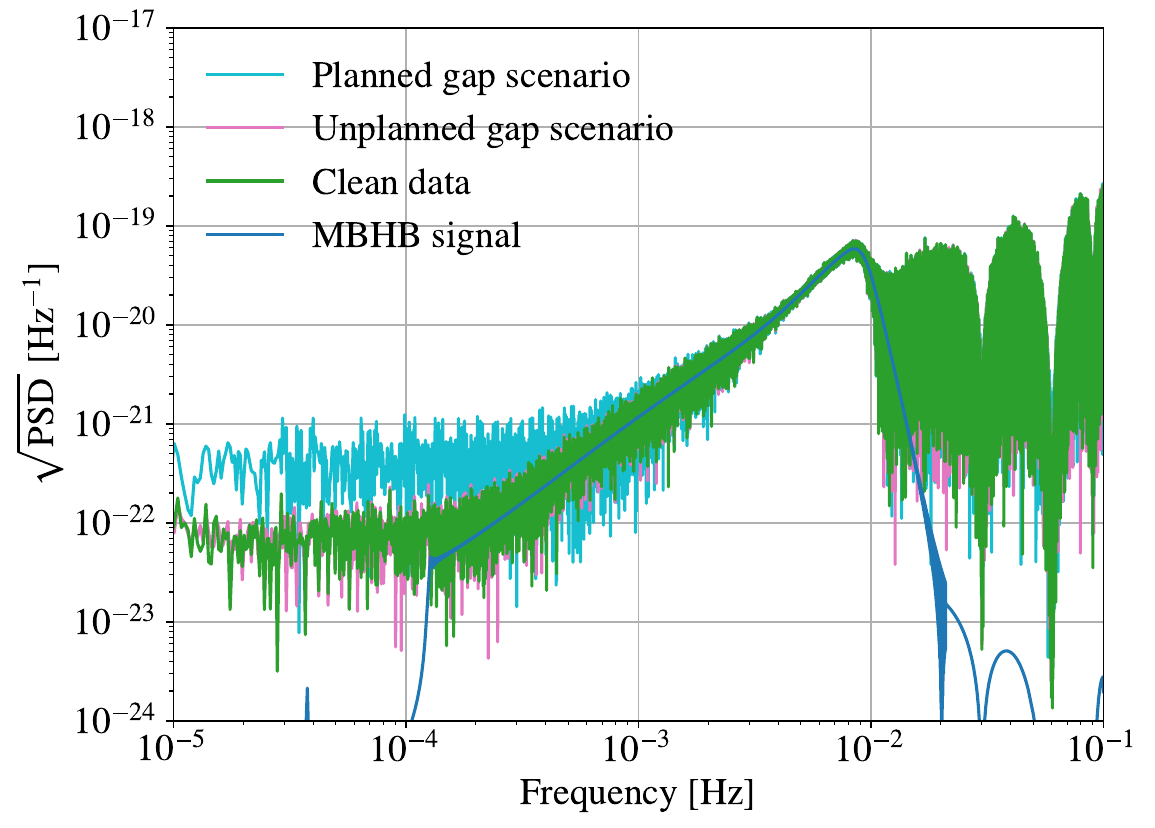}
    \caption{Comparison between the Amplitude Spectral Density $\sqrt{\text{PSD}}$ of data with different gap patterns applied to the \gls{MBHB} dataset. The green spectrum corresponds to the clean data (baseline without gaps). All periodograms are computed for a one-month \gls{TDI} channel $A$ time series, with no averaging.}
    \label{fig:noise-with-gaps-MBHB}
\end{figure}

In \Fref{fig:noise-with-gaps-MBHB}, we plot the Amplitude Spectral Density periodogram $\sqrt{\text{PSD}}$ of \gls{TDI} channel $A$ as a function of frequency $f$ for the \gls{MBHB} gap scenario, with no averaging window. 

The green spectrum corresponds to baseline data without gaps, includes Gaussian noise, and the underlying \gls{MBHB} signals are plotted in blue. The gapped scenarios represented on the plot are the planned gap scenario (cyan) and the unplanned gap scenario (pink), obtained by multiplying the clean data by the respective binary masks. Both gapped scenarios are affected by additional noise with respect to the baseline spectrum at frequencies lower than \SI{1}{mHz}, with the planned gaps scenario spectrum being the noisiest. This examination of the spectra suggests that the gap pattern under consideration impacts the inspiral phase of the \gls{MBHB} signal, with a more considerable impact from planned gaps than unplanned gaps.

\subsubsection{\label{sec:gap-mitigation-impact-on-mbhb-pe}Impact on parameter estimation}

\begin{figure}
    \centering
\includegraphics[width = 0.8\columnwidth]{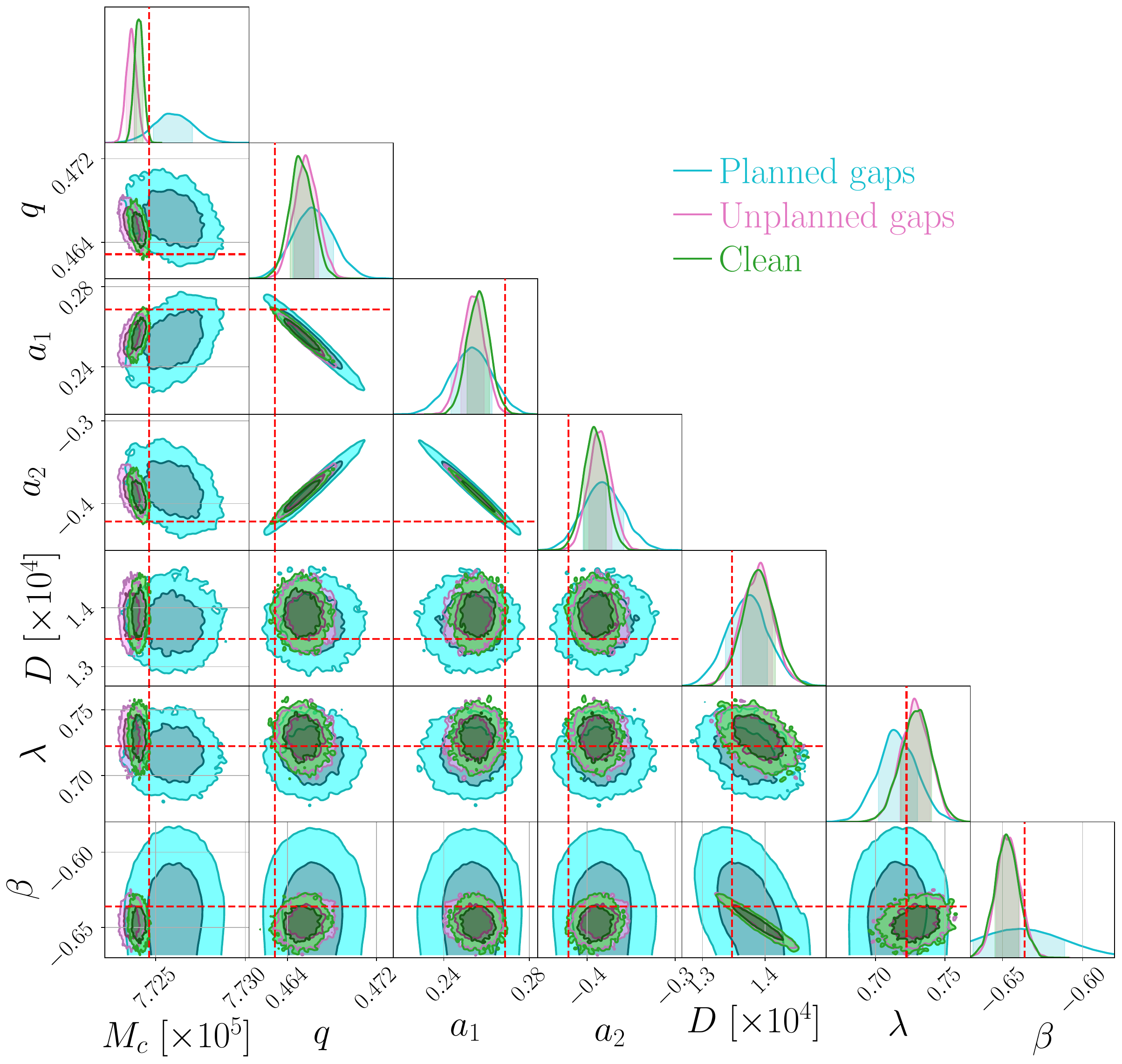}
    \caption{Corner plot of the \gls{MBHB} search parameters for the following datasets: clean data (green), planned gap scenario (cyan), unplanned gap scenario (pink). Depicted in the corner plots are the posterior probabilities for the following parameters: chirp mass $M_c$, mass ratio $q$, the dimensionless spin of the larger black hole $a_1$, the dimensionless spin of the smaller black hole $a_2$, luminosity distance $D$,  ecliptic longitude $\lambda$, ecliptic latitude $\beta$. Red dotted lines mark the true parameter values: {the \gls{MCMC} search was implemented with equal-armlength waveform approximation, while the \gls{LDC} data include an unequal-armlength LISA constellation \cite{spritz}, leading to a slight bias between the true parameter values and the clean data posterior.}}
    \label{fig:mbhb-posterior-gaps}
\end{figure}

We run a parameter estimation search for each gap scenario under consideration for the \gls{MBHB} dataset, using the same inference methods of Subsection~\ref{sec:glitch-mitigation-impact-on-mbhb-pe}. We obtain an estimate of the parameter posterior probability that we show in the corner plot of \Fref{fig:mbhb-posterior-gaps} including results for clean data (green), planned gap scenario (cyan), and unplanned gap scenario (pink), {alongside the injection values (red dashed lines)}. 

The presence of gaps induces both a widening of the posteriors, likely due to the consequence of \gls{SNR} loss, and a shift of the parameters mean due to the leakage in the signal. Indeed, the frequency-domain waveform used to estimate the parameters does not include the effect of data gaps.

Planned gaps have the largest impact on the estimated statistical uncertainty and the bias of the parameter posterior. They introduce a significant bias and an increase in the parameter estimation uncertainty of \glspl{MBHB} by a factor of 2 to 3 for all parameters under analysis. However, the parameter posteriors remain consistent with the injection, i.e. the estimated parameter values and their uncertainties are compatible with the source injected value. Nevertheless, it is paramount to plan data interruptions so that the time between predicted merger times and the nearest gap is as long as possible. This would ensure the maximization of the science return, as the 
bulk of the \gls{SNR} is concentrated around the merger.

Unplanned gaps have a minor impact on parameter estimation, though they show some bias and degraded uncertainty of about 20\% in the chirp mass $M_\text{c}$. {These results hold for the specific realization of the unplanned gap pattern under analysis, which includes two uncorrelated Poissonian environmental interruption 24 hour gaps, occurring about 20 and 7 days before merger. A different realization of the unplanned gap pattern, especially with gaps occurring in higher proximity to merger, could yield more impacting results.}  

In the considered scenario, gaps are sufficiently away from the \gls{MBHB} coalescence (or merger) to leave enough \gls{SNR} for detection and parameter estimation. The authors of \cite{Dey_2021} perform a study for two astrophysical distributions of \gls{MBHB}: one with heavy mass seeds and another with low mass seeds. The study addresses the detectability for each population based on \gls{FIM} computation. Bayesian parameter estimation is also performed on two representative sources drawn from each population. It was demonstrated that while planned gaps do not significantly impact \gls{MBHB} detectability, {preventing them from happening during the binary merger phase is crucial. Therefore, a good strategy allowing for protected periods is essential.} The authors also show that unplanned gaps have a statistically significant effect because of their longer duration, which increases the probability of the merger happening within one of them.

Naive mitigation approaches, like smooth windowing, are not applicable in the case of transient astrophysical sources, because they distort frequency-domain waveforms and reduce \gls{SNR} due to smooth transitions. 

\section{\label{sec:conclusions}Conclusions and perspectives}

We presented a summary of the analyses we carried out to prepare the LISA mission adoption concerning the robustness of current data analysis methods to data artifacts.

We investigated the impact of instrumental transient perturbations called glitches, considering \gls{LPF}-like force events acting on the \gls{LISA} \glspl{TM}. We show that completely ignoring their presence in the data analysis is incompatible with meeting the mission's science objectives, especially for continuous sources. The glitches' total power may completely dominate the \gls{GB} signal. 

{The shapelet model turned out to be a suitable mathematical description of the instrumental transients observed in LISA Pathfinder. They were particularly adapted to the sharp rise and exponential decays featured by the glitches. The model also offers flexibility when glitches exhibit more complex shapes, like double-sided rises. In that case, a linear combination of shapelet functions can be used to represent the glitches. In this work, we added the LISA instrumental response function to this model. To do so, we had to assume an injection point, i.e., a particular origin for the glitch process. To extend the generality of this description, more injection points different from TMs should be tested (leading to different responses).}

As an attempt to mitigate their impact, we use a method that first detects glitches by match-filtering the data using a shapelet model, and then applies a mask to discard the affected data in the analysis. This approach allows us to mitigate most of the impact on \gls{GB} parameter estimation, except for low-\gls{SNR} sources, for which more significant errors remain.

For transient \gls{GW} sources like \gls{MBHB}, glitches induce a bias and increase the estimated uncertainty in classic parameter estimation techniques. This calls for the necessity to account for their presence in the data analysis. As a countermeasure, masking yields acceptable results on \gls{MBHB} parameter estimation, with a few percent differences in posteriors.

We also investigated the impact of data gaps by considering two different patterns. One represents planned data interruptions, which are already anticipated today and over which we will have some control regarding occurrence times and duration through mission operations. Another pattern represents unplanned gaps, which may occur randomly in the data-taking process and for which no control is possible. As most of today's data analysis methods operate in the frequency domain, we show that the main impact of gaps comes from noise frequency leakage, inducing spectral distortions. If no treatment is applied, or if the frequency-domain covariance matrix is not accounting for them, these distortions may severely impact the characterization of continuous sources like \glspl{GB}. 
However, applying smooth windowing in the time domain with continuous transitions at the gap edges before the Fourier transform mitigates spectral leakage, provided that the smoothing is optimally chosen in terms of \gls{SNR}. It allows for accurate parameter estimation for most of the \gls{VGB} sources we consider. Again, sources close to the detection threshold are more impacted than others.

For the \gls{MBHB} case we studied, in the case of planned gaps we observe a significant bias and an increased parameter estimation uncertainty by a factor of 2 to 3. However, the parameter posteriors remain consistent with the injection, i.e.,  the estimated parameter values and their uncertainties are compatible with the source injected value. We obtain these results in a scenario where gaps arise far enough from the binary merger time. Nevertheless, it is paramount to plan data interruptions so that the time between predicted mergers and the nearest gap is as long as possible. This would ensure the maximization of the science return, as the bulk of the \gls{SNR} is concentrated around the merger. The \gls{SNR} may dramatically drop if this condition is not verified, down to the limit where the entire transient signal falls into the gap.

In the end mitigation of noise artifacts turns out to be a necessary step prior to parameter estimation. Our analysis shows that straightforward mitigation techniques can significantly mitigate artifacts, but residual impacts on the science remain. Future work on more sophisticated techniques can maximize the science return in the presence of imperfect data.

After this study, we identified several avenues of investigation. Regarding glitch detection, we point out the risk of confusion between glitches of instrumental origin and \gls{GW} transients or bursts. This impacts \gls{LISA}'s ability to detect \gls{GW} emissions from unmodeled sources, which is one of the mission's science objectives. Although not addressed in this work, techniques have been proposed to distinguish glitches and \glspl{GW} such as Bayesian model comparison based on sine-Gaussian wavelet models~\cite{Robson2019}, and applications of coronographic \gls{TDI} as a glitch veto~\cite{barroso2024coronagraphictimedelayinterferometrycharacterization}. Further developments are nevertheless needed to address this issue, accounting for the specificity of LISA glitch features, and the presence of other \gls{GW} sources. Modeling and fitting glitches as part of the parameter estimation can also avoid biasing parameter estimates~\cite{spadaro2023glitch}.

Besides, the treatment of data gaps may differ depending on the low-frequency noise power. In the model we use in this study, the second generation \gls{TDI} variables exhibit almost white noise below $10^{-4}$ Hz. If the noise were redder, leakage would be more important, calling for more sophisticated mitigation techniques than smooth masking. For example, \cite{Baghi_2019} shows that because of red noise, the effective \gls{SNR} can drop to 30\% of the result without gaps for low-frequency sources and frequent gaps. There are two options to cure this: statistical data imputation or accurate covariance modeling. In \cite{Baghi_2019} they opt for the former and develop a Bayesian data augmentation method that  reintroduces the missing data as auxiliary variables in sampling the posterior distribution of astrophysical parameters. Another inpainting algorithm inspired by sparse sensing and described in \cite{blelly2020recovery, blelly2021sparse} was  proposed to tackle this issue. Machine learning techniques like autoencoders are also being developed for data imputation, see \cite{mao_novel_2024}. Further research should be conducted to raise imputation methods to an operational level suitable to the global fit.

Finally, in this work, we assumed that the mechanisms at the origin of gaps and glitches do not modify the statistics of the Gaussian part of the noise. In other words, the Gaussian instrumental noise remains stationary and unperturbed throughout the time series. However, jumps in noise power could be triggered simultaneously with the glitch or gap processes. Such a situation would make the data analysis very different, requiring a noise model that differs between available (unperturbed) data segments. Changes in noise levels have already been considered in the context of stochastic \gls{GW} backgrounds in \cite{alvey2024}, and similar approaches should be tested for extracting other \gls{GW} sources.

\appendix
\section{\label{app:dfts}Analytical expressions for glitch waveform discrete Fourier transforms}
\setcounter{section}{1}

In this section we compute the \gls{DFT} of the glitch waveform from the time-domain model in Equations~\eqref{eq:phasemeter-glitch}-\eqref{eq:phasemeter-glitch-2}.

To ease the computation, we split the waveform in three components: a simple Heaviside function $v_{g1}(t) = H(t-t_0)$, an exponential decay $v_{g2}(t) = H(t-t_0) e^{\frac{(t-t_0)}{2\beta}}$ and a polynomial decay $v_{g3}(t) = H(t-t_0) e^{\frac{(t-t_0)}{2\beta}} \frac{t-t_0}{2 \beta}$, so that we have
\begin{equation}
    v_{g}(t) = \frac{2A}{\sqrt{\beta}} \left(v_{g1}(t) + v_{g2}(t)  + v_{g3}(t)\right).
\end{equation}
The DFT of the first component over a time series of size $N$ yields
\begin{equation}
     \tilde{v}_{g1}(f) = 
 \begin{cases}
 -n_0 & f = 0,\\
 e^{-i \pi f (n_0 + N -1)} \frac{\sin{\left(\pi f (N - n_0)\right)}}{\sin{\left(\pi f \right)}} & f \neq 0\\
 \end{cases}
\end{equation}
where $n_0 = \lfloor f_s t_0 \rfloor$.
The DFT of the second component yields
\begin{equation}
    \tilde{v}_{g2}(f) = \frac{1-e^{-i \left(\omega_p\right) (N-n_0)}}{1-e^{-i\left(\omega_p\right)}} e^{-i 2\pi f  n_0} e^{\frac{1}{2\beta}\left(t_0-\frac{n_0}{f_s}\right)},
\end{equation}
The third component writes:
\begin{align}
\tilde{v}_{g3}(f) & = \left(
    \frac{-\left(e^{i\omega_p}  \left(e^{-i\omega_pn}\left(n e^{i\omega_p}-(n-1)\right)-1\right)\right)}{2\beta f_s  (e^{i\omega_p}-1)^2} 
\right) \nonumber \\
& \times e^{-i 2 \pi f  n_0} e^{\frac{1}{2\beta}\left(t_0-\frac{n_0}{f_s}\right)}  - \left(t_0-\frac{n_0}{f_s}\right) \tilde{v}_{g2}(f),
\end{align}
where $\omega_p =  2 \pi f - \frac{i}{2\beta f_s}$ and $n = N - n_0$.

\ack
E.C.'s work is supported by NASA under award number 80GSFC24M0006. N.K. acknowledges the funding from the European Union’s Horizon 2020 research and innovation programme under the Marie Skłodowska-Curie grant agreement number 101065596. O.S.'s work is supported by the NASA LISA Preparatory Science program, grant number number 80NSSC19K0324. W.J.W. acknowledges support from the Istituto Nazionale di Fisica Nucleare (INFN) and Agenzia Spaziale Italiana (ASI), Project No. 2017-29-H.1-2020 “Attività per la fase A della missione LISA.”
\section*{References}
\bibliography{apssamp}

\end{document}